# Validating non-invasive EEG source imaging using optimal electrode configurations on a representative rat head model

Pedro A. Valdés-Hernández, Jihye Bae, Yinchen Song, Akira Sumiyoshi, Eduardo Aubert-Vázquez, Jorge J. Riera*

***Abstract***— **The curtain of technical limitations impeding rat multichannel non-invasive electroencephalography (EEG) has risen. Given the importance of this preclinical model, development and validation of EEG source imaging (ESI) is essential. We investigate the validity of well-known human ESI methodologies in rats which individual tissue geometries have been approximated by those extracted from an MRI template, leading also to imprecision in electrode localizations. With the half and fifth sensitivity volumes we determine both the theoretical minimum electrode separation for non-redundant scalp EEG measurements and the electrode sensitivity resolution, which vary over the scalp because of the head geometry. According to our results, electrodes should be at least ~3-3.5 mm apart for an optimal configuration. The sensitivity resolution is generally worse for electrodes at the boundaries of the scalp measured region, though, by analogy with human montages, concentrates the sensitivity enough to localize sources. Cramér-Rao lower bounds of source localization errors indicate it is theoretically possible to achieve ESI accuracy at the level of anatomical structures, such as the stimulus-specific somatosensory areas, using the template. More validation for this approximation is provided through the comparison between the template and the individual lead field matrices, for several rats. Finally, using well-accepted inverse methods, we demonstrate that somatosensory ESI is not only expected but also allows exploring unknown phenomena related to global sensory integration. Inheriting the advantages and pitfalls of human ESI, rat ESI will boost the understanding of brain pathophysiological mechanisms and the evaluation of ESI methodologies, new pharmacological treatments and ESI-based biomarkers.**

## I. ABBREVIATIONS

EEG: Electroencephalography
ESI: EEG Source Imaging
MRI: Magnetic Resonance Imaging (or Magnetic Resonance Image)
fMRI: Functional MRI

*Corresponding author: Jorge J. Riera is with the Florida International University, USA (e-mail: jrieradi@fiu.edu).

BEM: Boundary Element Method
FEM: Finite Element Method
LF: Lead field
AP: Anterior-posterior (rostro-caudal)
ML: Mediolateral
IS: Inferior-superior (ventro-dorsal)
HSV: Half Sensitive Volume
FSV: Fifth Sensitive Volume
CRLB: Cramér-Rao Lower Bound
RMS: Root-mean-square of the CRLB of the localization error
S1FL: Forelimb region of the primary somatosensory[1]
S1HL: Hindlimb region of the primary somatosensory
S1BF: Barrel field of the primary somatosensory
M1: Primary motor cortex
M2: Secondary motor cortex
Pt: Posterior parietal cortex
ERP: Event related potential
SEP: Somatosensory evoked potential (somatosensory ERP)
RGB: Red-Green-Blue color code

## II. INTRODUCTION

Electroencephalography (EEG) is presently an important tool for assisting diagnostics, management and treatment of neurological disorders. In particular, a better understanding of the electrical substrates of pathological events and abnormalities associated with brain network activity has been possible through the use of the estimated EEG sources within the brain, i.e. the EEG source imaging (ESI). Derived from standard EEG high density caps, ESI provides functional images of the whole brain with an exquisite temporal resolution. ESI has been used to determine important features of neuronal connectivity in autism spectrum disorders (Coben et al., 2014); to localize intrinsic sources of either interictal epileptiform discharges in focal epilepsy (Kaiboriboon et al., 2012) or those related to brain lesions (Harmony et al., 1995). It has also been useful in understanding important pathological mechanisms in psychiatric disorders (Hughes et al., 1999) and neurological conditions such as stroke (Bharadwaj y Gofton, 2015), attention deficit hyperactivity disorder (Rodrak y

---

[1] We'll use the anatomical structures of the Paxinos & Watson atlas (Paxinos y Watson, 2007)

Wongsawat, 2013) and Alzheimer's disease (Aghajani et al., 2013).

On the other hand, preclinical models are essential in studying most of these brain disorders. This is due to the possibility of combining invasive with non-invasive techniques, which promises of significant improvements to clinical treatments by establishing gold-standards, protocols and useful biomarkers. In particular, the rat is one of the most popular preclinical models, e.g. of Alzheimer's disease (Do Carmo y Cuello, 2013), epilepsy (Löscher, 2011), stroke (MacRae, 2011), autism (Umeda et al., 2010) and schizophrenia (Shevelkin et al., 2014). There is great interest from both research institutions and the pharmacological industry in developing practical tools for an efficacious evaluation of interventions using these preclinical models. Like in humans, rat ESI— provided that it is based on practical, feasible and exportable methodologies—would provide a promising non-invasive technique for chronic strategies that is needed to understand the evolution of the disease and the responses to specific treatments in this particular species.

Unfortunately, ESI in rodents has been challenging until recently due to the difficulties with obtaining whole scalp high density EEG recordings. Over the last decade, EEG recordings have been obtained with a few electrodes inserted inside the brain (Quairiaux et al., 2011). Even with numerous electrodes, the invasiveness of the methods used for electrode fixation poses an additional drawback, for they mostly employed surgical procedures to fix the electrode array, the electrodes are attached to the skull or under the skin (Choi et al., 2010; Franceschini et al., 2008; Lee et al., 2011; Mégevand et al., 2008). We have recently overcome this problem by developing the first EEG mini-cap to non-invasively record high-density (32-channels) scalp EEG in rats (Riera et al., 2012) (US patent Application No. 13/641,834). The methodology for using this mini-cap to perform ESI on a rat preclinical model of focal epilepsy was presented in Bae et al. (2015).

An additional, and very important, practical impediment for ESI is the achievement of realistic MRI-based forward models. This is prohibitive for large preclinical studies, for long EEG recordings precluding rat undergoing MRI, or when MRI scanners suitable for rats are not available. This very reason has already promoted the use of surrogate models in human ESI (Darvas et al., 2006; Valdés-Hernández et al., 2009). For rat ESI, there is a single precedent in Bae et al. (2015), where an approximate head model—with the consequent approximate electrode positions—was obtained from an MRI rat template. This template—rigorously called *minimum deformation template* since it minimizes the sum of the template-individual registration deformations across space and individuals—is considered as the most representative among individuals since it is the unbiased shape centroid of an MRI rat database (Valdés-Hernández et al., 2011).

Despite these important advances, rigorous theoretical analysis of the validity of the electrode density proposed by Riera et al. (2012) and of the accuracy in employing a template-based head model approximation is still missing. As a matter of fact, ESI would not be possible in rats if calculations from human data for optimal electrode resolutions (Gevins et al., 1990; Ferree et al., 2001; Malmivuo et al., 1997) and lower bounds for the dipolar localization errors (Mosher et al., 1993;

Beltrachini et al., 2011; von Ellenrieder et al. 2006) still hold for this particular species. Our goal in this paper is to validate practical tools used in the past by our group for ESI on rats by evaluating: 1) the electrode sensitivity resolution and 2) the lower bounds for the errors in the estimation of brain source introduced by the use of a template-based LF matrix. Errors from using this template have two sources: one related to the mismatches in the surfaces limiting different brain tissues and another because imprecisions in determining electrode positions in a head model.

Specifically, we firstly establish a reproducible rat EEG forward modelling methodology based on well-known numerical methods for calculating the LF matrix, i.e. the BEM and the FEM. We validate this methodology by investigating which is the optimal electrode configuration for scalp rat EEG. This is done by evaluating both the theoretical minimum distance at which two electrodes are not measuring redundant information and the electrode sensitivity resolution, using the concepts of half (HSV) and fifth (FSV) sensitivity volumes (Malmivuo et al., 1997; Wendel et al., 2008). Secondly, to theoretically validate the use of the template head model as a surrogate of an individual head model, we investigate if this model misspecification drastically increases the Cramér-Rao lower bound (CRLB) of EEG source estimation errors (Mosher et al., 1993) and provide, at the same time, the maximum possible accuracy of rat ESI. This is tackled with a method based on von Ellenrieder et al. (2006), which incorporates uncertainty in the LF matrix of the model, but deriving the covariance matrix directly from the LF matrices calculated from the individual MRIs of a rat database (individual LF matrices). Thereby this covariance accounts for both tissue-limiting surface mismatches and electrode location imprecisions. Since low CRLB values are only necessary conditions for an accurate ESI, we also compare the template LF matrix with the individual LF matrices using the Pearson product-moment correlation coefficient. This comparison is done for both BEM and FEM, as well as for different typical configurations of electrodes to account for the variability in electrode positioning of the rat EEG mini-cap. Finally we demonstrate the scope of our methodology using real ERP data obtained from Wistar rats. A first data set, which comprises simultaneous EEG and fMRI, is used to demonstrate both the accuracy of ESI and of the geometry-electrode template approximation. A second SEP dataset, with different sensory stimulation modalities, is used to illustrate the potential of the technique to extract information about traditional integration of sensory data at a whole brain global scale.

## III. MATERIAL AND METHODS

### A. Template set and individual MRIs

The details of the animal preparation, ethical considerations, MRI acquisition protocol and the template construction can be found in Valdés-Hernández et al., (2011). Briefly, we acquired the MRI of $N_K$=28 Wistar rats (Charles River Japan, Yokohama, Japan). They consist of respiratory-gated 2D TurboRARE T2 images with parameters: TR/TE$_{eff}$=10971/30, RARE factor=4, effective SBW=100 kHz, FOV=32×32 mm², matrix size = 256×256, in-plane resolution=125×125 μm², slices=128 and slice thickness=0.3 mm. The template set was built using these individual MRIs and comprises 1) an unbiased average of the nonlinearly warped MRIs, 2) probabilistic maps of brain, gray matter, white matter and cerebrospinal fluid, and 3) a discrete image labeling 96 structures of the cortex, according to the Paxinos and Watson classification atlas (Paxinos y Watson, 2007).

### B. Forward problem of the EEG

The estimation of the brain electrical sources of EEG is known as the *EEG inverse problem*. For solving this problem, the *EEG forward problem* must be defined. The solution of this problem provides the voltage generated at the EEG electrodes by any bioelectrical generator inside the brain. Although recent revolutionary viewpoints reconsider the role of electrical monopoles (Riera et al., 2012), there is still a consensus to model a *generator* of the EEG as a current dipole, that is, a *source* (-I) and a *sink* (I) electrical currents separated by a distance d, which is small compared to the distance between the dipole and the electrode. The amplitude of the dipole is Q=Id. For the used notation see Appendix A.

For one time instant, let us build the vector comprising $N_D$ dipoles $\mathbf{j} \equiv \begin{bmatrix} \mathbf{j}_1^T & \dots & \mathbf{j}_{N_S}^T \end{bmatrix}^T \in \mathbb{R}^{3N_D \times 1}$, where $\mathbf{j}_j = Q_j \mathbf{d}_j$ is the $j-th$ dipole with orientation $\mathbf{d}_j \in \mathbb{R}^{3\times1}$. Given the position of $N_E$ scalp EEG electrodes, the forward problem reduces to calculating the LF matrix, which relates these bioelectrical generators to the voltage differences $\mathbf{v} \in \mathbb{R}^{N_E \times 1}$ they create at the electrodes with respect to a common reference:

$$\mathbf{v} = \mathbf{K}\mathbf{j} + \boldsymbol{\varepsilon} \qquad (1)$$

where $\mathbf{K} \in \mathbb{R}^{N_E \times 3N_D}$ is the LF matrix, defined as

$$\mathbf{K} = \begin{bmatrix} \mathbf{k}_{1,1} & \mathbf{k}_{1,2} & \dots & \mathbf{k}_{1,N_D} \\ \mathbf{k}_{2,1} & \mathbf{k}_{2,2} & \dots & \mathbf{k}_{2,N_D} \\ \vdots & \vdots & \ddots & \vdots \\ \mathbf{k}_{N_E,1} & \mathbf{k}_{N_E,2} & \dots & \mathbf{k}_{N_E,N_D} \end{bmatrix} \qquad (2)$$

where the elements of the vector $\mathbf{k}_{i,j} \in \mathbb{R}^{1\times3}$ represent the voltage difference associated with the $i-th$ electrode of an unit amplitude dipole at the $j-th$ location, oriented along the X, Y and Z axes, respectively. The vector $\boldsymbol{\varepsilon} \sim N\left(\mathbf{0}_{N_E \times 1}, \sigma^2 \mathbf{I}_{N_E}\right)$ is a multivariate Gaussian white noise.

We calculate the LF matrix using two numerical methods:

1. The BEM, using our own MATLAB code (Riera y Fuentes, 1998).

2. The FEM, using SimBio, a generic environment for bio-numerical simulations from the SimBio Development Group[2].

### C. Head modelling

For calculating the LF matrix it is necessary to physically model the head. This requires a model of the possible locations and orientations of current dipoles within the brain (i.e. *the*

*source or dipole space model*) and the distribution of electrical conductivities in the head to simulate secondary conduction currents (i.e. *the volume conductor model*). To build a realistic head model, an individual MRI of the rat, or an equivalent representative (i.e. the template MRI), is required to extract the different compartments defining both the source and the volume conductor models.

#### 1) Source space models

We considered two types of source space models, which are shown in Figure 1A:

1. *Volumetric model*: with 3D support either inside the whole brain or only the gray matter. These were discretized into regular grids of points equally spaced along each direction, considered as the possible dipole locations. Since their orientations were left unconstrained, the LF matrix for each grid point was calculated along X, Y and Z axis and takes the full form of Eq. (2).

2. *Surface model*: defined as the surface amidst the pial surface and the boundary between the cortical layer and the white matter. This was discretized to a triangular mesh to define the possible dipole locations. Their orientations were constrained along the normal to the surface, based on the hypothesis that primary currents parallel to the pyramidal neurons, which are oriented perpendicular to the cortical layers, are the main cause of the rat scalp EEG (Riera et al., 2012). This means that all $\mathbf{k}_{i,j}$ vector elements in Eq. (2) are substituted by the scalar $k_{i,j} = \mathbf{k}_{i,j}^T \cdot \hat{\mathbf{n}}_j$, where $\hat{\mathbf{n}}_j$ is the normal to the surface at vertex $j$.

In order to compare the LF matrix calculated using the individual MRI (individual LF matrix) for each rat and the respective representative LF matrices calculated using the template (template LF matrix), the points of the discretization of the source space models in these two MRI-based coordinate systems must be equivalent. To achieve this one-to-one *anatomical correspondence* we generated a unique discretization in the template and transformed the points to all the individual spaces. This is related to the notion of the "canonical mesh" used in SPM8 (Litvak y Friston, 2008; Mattout et al., 2007)[3], rigorously applied for surfaces in Valdés-Hernández et al. (2009). Since the LF matrix comparisons were only performed for volumetric models, the transformations were estimated by spatially normalizing the individual MRIs to our MRI template using SPM8 "unified segmentation"[4]. This procedure optimizes a cost function based on MRI intensity similarity, instead of source space feature similarity. Therefore a tiny amount of transformed grid points fell outside of each individual brain and were discarded in both individual and template volumetric source models.

---

*2) Volume conductor model*

Using FreeSurfer[5], we divided the template and all individual MRIs into three triangular meshes defining the boundaries of three compartments of homogeneous conductivity representing the brain, the skull and the skin, with 0.33 S/m, 0.0041 S/m and 0.33 S/m, respectively. Figure 1A depicts this delineation.

For BEM (Riera y Fuentes, 1998), the Delaunay triangulation method was used to parametrically represent these surfaces with 2432 vertices (see Figure 1B). For FEM, these BEM compartments were divided into meshes of tetrahedra using Tetgen 1.50. The shape of these tetrahedra was constrained by the quality criterion q=1.414 which is the ratio of the radius of the circumsphere to the shortest edge of the tetrahedron. Volumes were constrained to be no larger than $1mm^3$, $0.5mm^3$ and $0.5mm^3$ for the skin, the skull and the brain, respectively. This yielded a mesh of about 4 million tetrahedra. Figure 1C shows the tetrahedral meshes.

The FEM method used requires the fulfilment of the Venant's condition (Vorwerk et al., 2012), which restricts source points to be in a node of the mesh with all neighboring elements belonging to the same conductivity compartment. To guarantee this without shifting the source positions and compromising the anatomical correspondence between source spaces, we forced the source points to be nodes of the tetrahedra during the meshing procedure.

Both BEM and FEM methods actually calculates the LF matrix at the triangle vertices defining the scalp surface. In the former case, interpolation furnished the LF matrix values at the electrodes positions. In the latter, we also forced the electrode positions to be nodes of the tetrahedral mesh.

The Table 1 shows the head models built for this paper.

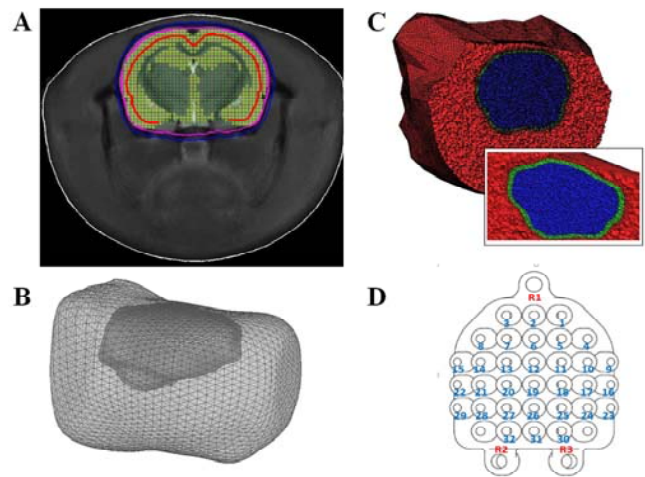

Fig. 1. **Head models of the rat**. A) The delineation of the three conductivity compartments is shown in an axial slice: head (white), skull (blue) and brain (magenta). The source space grid points of model Vol3D and VolGM3D are shown as green dots and yellow circles respectively. The intersection between the source model of SurfMid and the slice is shown with a red curve. B) Delaunay BEM triangulations. C) Tetrahedral FEM mesh. D) Electrode EEG mini-cap scaffold. The electrodes are labelled with blue numbers and the landmarks used to aid placing the cap in the MRI are labelled in red.

Table 1. **Different head models used to define the EEG forward problem**

| Model | MRI | Source Space | # Dipole Locations | Volume Conductor Modeling | Observations |
|---|---|---|---|---|---|
| **VolB** | • 28 individual rats of the database<br>• Template. | • Volumetric<br>• Whole brain<br>• Unconstrained dipole orientation | ~43k | BEM / FEM | Whole brain comprises white matter, gray matter and cerebrospinal fluid. Grid resolution along X, Y and Z is 0.35 mm. This model is used for the individual-template comparisons, the evaluation of electrode resolution and the estimation of the Cramér-Rao Lower Bounds (CRLB). |
| **VolGM** | • Template<br>• Rat of the EEG-fMRI experiment | • Volumetric<br>• Gray matter<br>• Unconstrained dipole orientation | ~28k | FEM | It is a subset of the previous under the hypothesis that EEG sources are in the gray matter. Restricted to the maximum probability mask of gray matter, created from the Template Set tissue priors. Used for visualization of the CRLB results and as a source space model for solving the inverse problems for real data. |
| **SurfMid** | • Template<br>• Rat of the EEG-fMRI Experiment | • Surface<br>• Mid cortical layer<br>• Dipoles constrained normal to the cerebral cortex | ~9500 | FEM | Surface amidst the pial and the gray/white matter interface. Used as a source space model for solving the inverse problems for real data. |

### D. Electrode sensitivity and resolution

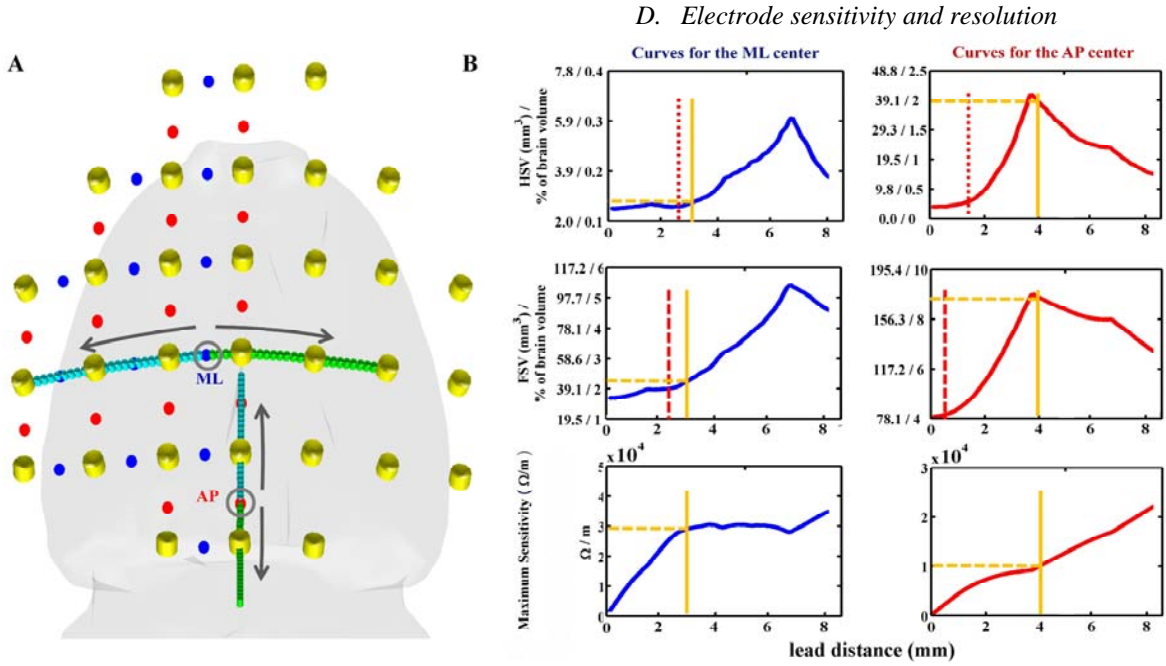

Fig.2.   **Sensitivity analysis of the rat EEG mini-cap.** A) A schematic of the position of the centers chosen for calculating the maximum scalar sensitivities, the HSV and FSV for increasing separating leads. The separation is indicated with the gray arrows along the cyan and green cylinders to both sides of the centers. Blue and red dots represent the centers for leads along the ML and AP directions, respectively. The centers were chosen as the middle points between each pair of electrodes of the mini-cap (represented as yellow cylinders). B) Curves of S, HSV and FSV versus geodesic distance for two examples. The HSV and FSV are also reported in fractions of the brain volume. In all six curves, the yellow vertical line indicates the electrode separation of the mini-cap (3 mm for ML and 4 mm for AP). In the HSV and FSV curves, the red broken vertical line indicates the maximum spatial electrode resolution.

#### 3) Electrode positioning

The EEG mini-cap scaffold contains 32 electrodes and three jutting circles used as landmarks (Bae et al., 2015), as shown in Figure 1D. After identifying the latter in the MRI template, the electrode positions are generated in the corresponding head model. The details about this procedure can be found in Bae et al. (2015).

In this paper we generated the electrodes positions on the scalp template using this strategy. Based on our experience, we randomly shifted the landmarks to simulate a small lack of precision in the routinely experimental procedure to place the mini-cap on the rat head. The larger shifts were conferred along the AP direction (maximum ~2mm) since an experimentalist is less prone to shift the mini-cap to one side of the rat head given its lateral symmetry. Thereby, we produced five different mini-cap configurations. We located the electrodes in each individual of the MRI in a different manner. We first took a photo of the electrodes in the template. Several volunteers, familiarized with both rat anatomy and EEG recordings, were asked to take this photo as a guide to manually place the mini-cap on each individual scalp surface model extracted from MRI the dataset, again using the landmark-based strategy. This second procedure simulates the errors committed by a researcher trying to identify the electrode locations during head modeling based on the experimental documentation. This was done for the five configurations of electrodes.

A *lead* is a common concept in EEG studies. It is a pair of electrodes on the scalp, say $i_E$ and $i_R$. Given a current directly applied to the *lead*, the current density distribution generated inside the volume conductor is proportional to $\mathbf{k}_{i_E,j} - \mathbf{k}_{i_R,j}$, for every source point $j$. The Helmholtz's reciprocity theorem reverses this view stating that this vector provides the sensitivity distribution of the lead (Malmivuo et al., 1997; Rush y Driscoll, 1969). The scalar sensitivity for a lead is defined as $S = \left\| \mathbf{k}_{i_E,j} - \mathbf{k}_{i_R,j} \right\|_2$.

It is desirable to have a measurement sensitivity as concentrated in a small volume as possible, because localization of sources is not possible with leads which sensitivity is homogenous over the entire source space. The concept of HSV is used to quantify the ability of the lead to concentrate its sensitivity in a region (Malmivuo et al., 1997). It is the volume of the source region for which the scalar sensitivity is higher than half of the maximum scalar sensitivity. Likewise, the FSV (Wendel et al., 2008) is the volume of the source region for which the scalar sensitivity is higher than a fifth of the maximum scalar sensitivity. The greater the HSV and the FSV, the more homogenous the sensitivity profile and the worse the ability of the lead to accurately localize sources. Thus, the HSV and FSV provide measures of electrode resolution.

The HSV and FSV have been evaluated in spherical models of human head sizes (Ferree et al., 2001; Malmivuo et al., 1997;

Malmivuo y Suihko, 2004; Wendel et al., 2008), which is a simplified but good approximation of the human head geometry. The common practice is to calculate curves of HSV/FSV versus angle or distance separation of the lead. From these curves conclusions regarding electrode resolution of different EEG caps are drawn. The HSV/FSV values for specific lead separations correspond to the resolution of certain electrode cap densities. For example, for a scalp sphere of radius $R = 9.2 cm$, the separations $12.4°$ ( $2cm$ ), $15.5°$ ( $2.5cm$ ) and $31°$ ( $5cm$ ) corresponds to the human montages of 256, 128 and 32 electrodes, respectively. Besides, the limits for EEG cap density can be identified from the HSV/FSV curves. For increasing cap density (decreasing lead separation) the curve decreases until a baseline is reached. This baseline is the maximum reachable sensitivity resolution and the separation at which it starts was called the *maximum spatial electrode resolution* (Ferree et al., 2001). Electrode caps with densities such that the electrode separation is below the maximum spatial electrode resolution are measuring redundant information.

Due to the rotational invariance of the spherical model, the center of the separating leads can be placed at any point in the scalp. Thus the conclusions drawn from these previous curves are valid for every pair of electrodes in an EEG cap. However, the situation complicates for realistic geometries. Since there is no rotational invariance, we expect the HSV/FSV curves to depend on the location of the lead center. As mentioned above, the spherical models may be enough to draw useful conclusions about human settings. In contrast, the rat head is not well approximated by concentric spheres (see Figure 1B). The scalp is cylindrical-like in the AP direction, yet the skull and the brain are ellipsoidal. This implies that the skull-scalp separation is highly variable, steeply increasing along the ML direction. This suggests that the HSV/FSV-based electrode resolution analysis in scalp EEG of rats is scalp position-dependent. We envisaged that different conclusions should be drawn for every pair of electrodes in the rat EEG mini-cap.

To be exhaustive, we calculated the HSV/FSV curves for centers located between every pair of electrodes of the rat EEG mini-cap in both the AP and the ML directions. From each curve, we quantified two measures of the sensitivity resolution of the electrode pair. These are the HSV and FSV values in the curves at the separation of the electrode pair. On the other hand, to ensure the electrode pair is not measuring redundant information, we checked if its separation is above the maximum spatial electrode resolution, also identified from the HSV and FSV curves. Since we can derive different maximum electrode resolution for each curve type, we take the maximum among both. The sensitivity was obtained from the LF matrix of the VolB model (Table 1) and the consequent HSV/FSV values were calculated by summing up the number of grid points and multiplying by the grid voxel size. To build the curves, the lead separations ranged from $0.05 mm$ to 15 mm at every 0.05 mm. Figure 2 illustrates how the whole procedure was done.

In spherical models, for small lead separations, the sensitivity profile within the source region contributing to the HSV is predominantly tangential to the electrodes (Malmivuo et al., 1997), though there are still contributions from radial dipoles around the source space points closest to the electrodes (Ferree et al., 2001). For increasing lead separation the radial contribution increases while the tangential decreases until, for very well separated electrodes, the HSV region is split in two identical sub-regions. Each sub-region surrounds the source space below each electrode, with the complete radial sensitivity profile described for one electrode (Ferree et al., 2001; Malmivuo y Suihko, 1997; Malmivuo y Suihko, 2004).

For the realistic rat geometry, this might not be straightforward. Thus, we also investigated the sensitivity to the main dipole orientation of the rat EEG mini-cap. For summarizing results in a simple manner, we propose a single vector that averages the orientation sensitivity profile across all the points within the HSV region. Given the sensitivity of a given electrode pair, i.e. $\mathbf{s}_j = \mathbf{k}_{1g,j} - \mathbf{k}_{1g,j}$, for which we omit the electrode indices for simplicity, let's define the matrix:

$$\mathbf{C} = \frac{1}{N_{HSV}} \sum_{j \in HSV} \mathbf{s}_j \mathbf{s}_j^T, \quad (3)$$

where $N_{HSV}$ is the number of source points within the HSV region. Let $\lambda_i$ and $\mathbf{v}_i$, for $i = 1, 2, 3$, be the eigenvalues and eigenvectors of $\mathbf{C}$ ( $\lambda_1 > \lambda_2 > \lambda_3$ ). Since $\mathbf{C}$ is positive-definite and symmetric, its eigenvalues are positive and real. Therefore, $\mathbf{C}$ can be geometrically represented as an ellipsoid with principal axis $\mathbf{v}_1$ denoting the main orientation of the sensitivity profile within the HSV.

For non-degenerate ellipsoids eccentricity of the ellipsoid takes values in the interval $[0,1)$ and it is given by:

$$ecc = \sqrt{\frac{3\sum_i (\lambda_i - \bar{\lambda})^2}{2\sum_i (\lambda_i)^2}}, \quad \bar{\lambda} = \frac{\lambda_1 + \lambda_2 + \lambda_3}{3} \quad (4)$$

For $ecc = 0$, the distribution of the sensitivity is isotropic, while for $ecc \to 1$ the sensitivity is anisotropic. Helmholtz's theorem allows us to view the sensitivity of a lead as currents flowing from one electrode to the other. Provided that the conductivities are homogeneous inside the compartments and there is only one source and one sink, which are the electrodes, the LF cannot be isotropic at any region of the source space. Therefore we do not expect low eccentricity values. However, even with a high eccentricity, the ellipsoid can be either "prolate" ( $\lambda_1 \gg \lambda_2 \approx \lambda_3$ ) or "oblate" ( $\lambda_1 \approx \lambda_2 \gg \lambda_3$ ). In the former case, it is clear that the sensitivity profile is mainly toward the principal eigenvector. However, for the latter case the orientation distribution is shared between the first two eigenvectors. To account for this, we propose the following vector to quantify the main orientations in the sensitivity distribution and also accounts for its variability:

$$\mathbf{c} = ecc \frac{(\lambda_1 \mathbf{v}_1 + \lambda_2 \mathbf{v}_2)}{\lambda_1 + \lambda_2} \quad (5)$$

For prolate ellipsoids $\mathbf{c}$ is approximately collinear with $\mathbf{v}_1$ while for oblate ellipsoids is the weighted average of the first two principal directions. The size of this vector, given by ecc, denotes orientation coherence. We calculated $\mathbf{c}$ for every ML and AP electrode pair of the EEG mini-cap.

## E. Source estimation errors. The Cramér-Rao Lower Bound

Irrespective of the estimation method, the Cramér-Rao theorem provides necessary conditions for the accuracy of the estimation of any vector of parameters $\boldsymbol{\theta}$. It establishes the CRLB for the covariance of the difference between an unbiased estimator and the actual parameter values, i.e.:

$$\mathrm{E}\left\{\left(\boldsymbol{\theta}-\hat{\boldsymbol{\theta}}\right)\left(\boldsymbol{\theta}-\hat{\boldsymbol{\theta}}\right)^T\right\} \geq \mathbf{F}^{-1}$$

$$\mathbf{F} = \mathrm{E}\left\{\left[\frac{\partial}{\partial \boldsymbol{\theta}}\log p\left(\mathbf{v} \mid \boldsymbol{\theta}\right)\right]\left[\frac{\partial}{\partial \boldsymbol{\theta}}\log p\left(\mathbf{v} \mid \boldsymbol{\theta}\right)\right]^T\right\} \quad (6)$$

where $\mathbf{F}$ is the Fisher information matrix, $\mathbf{v}$ is the voltage and $\mathrm{E}\{\cdot\}$ denotes expectation.

Using the LF matrix of a spherical model fitted to the human head size, the Cramér-Rao theorem was firstly applied in Mosher et al. (1993) for assessing the CRLB of the estimation of the positions of dipoles given the EEG forward problem of the type in Eq. (1), but accounting for several time instants. This general spatiotemporal model is presented in the Eq. (B1) in Appendix B.

For this paper, we applied this theory using the rat LF matrix. The vector of estimated parameters is $\boldsymbol{\theta} = \begin{bmatrix} \sigma^2 & \bar{\mathbf{J}}^T & \mathbf{r}^T \end{bmatrix}^T$, where $\bar{\mathbf{J}} = \begin{bmatrix} \mathbf{j}^T(1) & \dots & \mathbf{j}^T(N_T) \end{bmatrix}^T$ is the grand vector concatenating all $N_D$ dipoles $\mathbf{j}(t)$ at all time-points $t = 1 \dots N_T$. The vector $\mathbf{j}(t)$ is the same used in Eq. (1) but now with temporal dependency. The vector $\mathbf{r} \in \mathbb{R}^{3N_D \times 1}$, with $\mathbf{r} = \begin{bmatrix} \mathbf{r}_1^T & \dots & \mathbf{r}_{N_S}^T \end{bmatrix}^T$ and $\mathbf{r}_k = \begin{bmatrix} r_{kx} & r_{ky} & r_{kz} \end{bmatrix}^T$, contains the position of the dipoles.

In Mosher's analysis, the only cause of error is the observational noise encoded in the parameter $\sigma$. Based on the work of von Ellenrieder et al. (2006) we also considered an additional cause of error: that of using the template LF matrix as a surrogate of the individual LF matrix. This error accounts for both geometry and electrode misspecifications. In this case the LF matrix is written as $\mathbf{K} = \mathbf{K}^{(0)} + \Delta \mathbf{K}$ where $\mathbf{K}^{(0)}$ is the template LF matrix and $\Delta \mathbf{K}$ is a random variable which encodes the variability of the LF matrix across different rats. If the parameters that contribute to the calculation of $\mathbf{K}$ (e.g. shape of the surfaces, conductivities, etc.) have a normal variability and the variations are small enough, $\Delta \mathbf{K}$ is also multivariate normal (von Ellenrieder et al., 2006). We expected small variations in our case given the usual similarity of the rats. This can be in fact evidenced by the very high values of the Pearson correlation coefficients between the individual LF matrices and the template LF matrix (see Section I.E and II.C). To ensure the approximation, we directly performed an element-wise Kolmogorov-Smirnov test under the null hypothesis of zero-mean normality.

Since both observational noise and $\mathbf{K}$ are normal, the voltage is also normal and the Fisher information matrix is calculated using (Kay, 1993):

$$F_{pq} = \frac{\partial \bar{\mathbf{V}}}{\partial \theta_p} \mathbf{C}_V^{-1} \frac{\partial \bar{\mathbf{V}}}{\partial \theta_q} + \frac{1}{2} tr\left\{\mathbf{C}_V^{-1} \frac{\partial \mathbf{C}_V}{\partial \theta_p} \mathbf{C}_V^{-1} \frac{\partial \mathbf{C}_V}{\partial \theta_q}\right\}, \quad (7)$$

where the grand vector $\bar{\mathbf{V}} = \begin{bmatrix} \mathbf{v}^T(1) & \dots & \mathbf{v}^T(N_T) \end{bmatrix}^T$ comprises the voltage vector for the $N_T$ time points. Its covariance matrix is:

$$\mathbf{C}_V = \mathrm{E}\left\{\Delta \mathbf{G} \, \bar{\mathbf{J}} \bar{\mathbf{J}}^T \, \Delta \mathbf{G}^T\right\} + \sigma^2 \mathbf{I}_{N_E N_T}$$

$$\Delta \mathbf{G} = \mathbf{I}_{N_T} \otimes \Delta \mathbf{K} \quad (8)$$

To calculate (7), we proposed the use sample estimators across all $N_K$ individuals and the 5 electrode configurations mentioned above, i.e.:

$$\hat{\mathbf{C}}_V = \frac{1}{5N_K - 1}\sum_{e}^{5}\sum_{k}^{N_K} \Delta \mathbf{G}^{(k,e)} \, \bar{\mathbf{J}} \bar{\mathbf{J}}^T \left(\Delta \mathbf{G}^{(k,e)}\right)^T + \sigma^2 \mathbf{I}_{N_E \cdot N_T}$$

$$\widehat{\frac{\partial \mathbf{C}_V}{\partial \theta_p}} = \frac{\partial}{\partial \theta_p}\left(\hat{\mathbf{C}}_V\right) \quad (9)$$

$$\Delta \mathbf{G}^{(k,e)} = \mathbf{I}_{N_T} \otimes \left(\mathbf{K}^{(k,e)} - \mathbf{K}^{(0,e)}\right)$$

$$k = 1 \dots N_K$$

We used the VolB model (see Table 1) to calculate $\mathbf{K}^{(0,e)}$, $\mathbf{K}^{(k,e)}$ and their derivatives, for $k = 1 \dots N_K$ and $e = 1, 2, 3, 4, 5$.

The Appendix B provides the detailed derivation of the estimation of the CRLB for the errors of estimated dipole locations and amplitudes. These, denoted by $CRLB(\mathbf{r})$ and $\mathbf{CRLB}(\bar{\mathbf{J}})$, are calculate by using Eqs. (B10) and (B11), which we rewrite here:

$$\mathbf{CRLB}(\mathbf{r}) = \left(\mathbf{F}_{rr} - \mathbf{F}_{rj}\mathbf{F}_{jj}^{-1}\mathbf{F}_{rj}^T\right)^{-1}$$

$$\mathbf{CRLB}(\bar{\mathbf{J}}) = \left(\mathbf{F}_{jj} - \mathbf{F}_{rj}^T\mathbf{F}_{rr}^{-1}\mathbf{F}_{rj}\right)^{-1} \quad (10)$$

where $\mathbf{F}_{rr}$, $\mathbf{F}_{jj}$ and $\mathbf{F}_{rj}$ are sub-matrices of the Fisher information matrix, given by Eq. (B6). The sub-indices "r" and "j" are related to the derivatives in Eq. (7) with respect to the dipole position and amplitude.

This formulation allows for the estimation of the CRLB for any number, orientation, amplitude and temporal behavior of the dipoles in the source space. Analyzing all combinations is impossible. This is the reason we considered, like in Mosher et al. (1993), only two situations a) one dipole ($N_D = 1$) and b) two dipoles ($N_D = 2$) of the same amplitude (but different orientations and locations), both cases at a single time instant. The diagonal elements of the matrices $\mathbf{CRLB}(\mathbf{r})$ and $\mathbf{CRLB}(\bar{\mathbf{J}})$ are the lower bounds of the variances of both errors. If $\gamma_{ij}$ are the entries of $CRLB(\mathbf{r})$, a scalar error bound measure is defined as the root-mean-square of the standard deviations of the dipole locations (Mosher et al., 1993). For both the single and double dipole cases:

$$RMS_1 \equiv \sqrt{\gamma_{11} + \gamma_{22} + \gamma_{33}} \quad \text{for dipole 1}$$
$$RMS_2 \equiv \sqrt{\gamma_{44} + \gamma_{55} + \gamma_{66}} \quad \text{for dipole 2} \quad (11)$$

where "dipole 1" accounts for either the single dipole or one of the dipoles in the double dipole case. If $\beta_{ij}$ are the entries of $\mathbf{CRLB}(\mathbf{J})$, analogous measures can be also defined for the amplitude. However, we find more informative to define a standardized measure:

$$Z_1 \equiv \frac{Q_1}{\sqrt{\beta_{11} + \beta_{22} + \beta_{33}}} \quad \text{for dipole 1}$$

$$Z_2 \equiv \frac{Q_2}{\sqrt{\beta_{44} + \beta_{55} + \beta_{66}}} \quad \text{for dipole 2}$$

(12)

If $Z_{1,2} \leq 1$ the estimated amplitude is smaller than its error and it is therefore not significant. We deem the scalars in (12) as quantitative measures of the *upper bounds* for the "estimability" or "detectability" of the dipoles.

For the single dipole case, we calculated the RMS and Z for all dipoles oriented along X, Y or Z in the gray matter region defined as the source space model of VolGM (Table 1). For the double dipole case, we fixed the position and orientation of one dipole which we call the "fixed" dipole and calculated RMS and Z of both dipoles for every possible position of the second dipole, which we shall call the "moving" dipole. Both were also restricted to the VolGM source space. Although this restriction was mainly conceived for saving computational cost, it is actually a physiological constraint for the dipole locations. Remember, however, that the derivatives of the LF matrices were calculated using VolB to avoid undesirable boundary effects.

As mentioned above, we conferred, for simplicity, the same amplitude to all dipoles, i.e. $\mathbf{j}_k = Q\hat{\mathbf{d}}_k$, where $\hat{\mathbf{d}}_k$ is the unit vector pointing along the dipole. In this case, the sub-matrices $\mathbf{F}_{rr}$ and $\mathbf{F}_{rj}$ in Eq. (10) are proportional to $Q^2$ and $Q$, respectively. This implies that $\mathbf{CRLB}(\mathbf{r})$ and $\mathbf{CRLB}(\mathbf{j})$ are proportional to $Q^{-2}$ and $Q^2$, respectively. Then $Q$ factors out of the square root in the denominator of Z in Eq. (12), cancelling out with the $Q$ in the numerator. In general, we state without providing a trivial demonstration that the assumption of equal amplitudes allows for the substitutions $\sigma \to \sigma/Q$ and $\mathbf{j}_k \to \hat{\mathbf{d}}_k$ in the whole formulation given in the Appendix B. Now both the RMS and Z only depend on the LF matrices, the factor $f = \sigma/Q$ and the positions and orientation of the dipoles. On the other hand, as expected, in the absence of LF matrix errors, i.e. $\Delta\mathbf{K} = 0$, the formula to obtain the RMS is identical to that derived by Mosher et al. (1993) and linearly proportional to $f$, while Z is inversely proportional to $f$.

### F. Comparison of the LF matrices

We compared the LF matrices using the Pearson product-moment correlation coefficient:

$$\rho_j^{(k)} = \frac{\left\langle \mathbf{K}_j^{(k)}, \mathbf{K}_j^{(0)} \right\rangle_F}{\left\| \mathbf{K}_j^{(k)} \right\|_F \left\| \mathbf{K}_j^{(0)} \right\|_F}$$

(13)

The sub-array $\mathbf{K}_j = \begin{bmatrix} \mathbf{k}_{1,j}^T & \dots & \mathbf{k}_{N_E,j}^T \end{bmatrix}^T$ is the LF matrix of the $j$-$th$ source for both the individual and template LF matrices, denoted by $\mathbf{K}^{(k)}$, k=1,…,$N_K$ and $\mathbf{K}^{(0)}$, respectively. The symbol $..$ represents inner product under the Frobenius norm $\|\cdot\|_F$. This correlation coefficient takes values in the interval [-1,1] where $\rho = 1$ represents full correlation, $\rho = -1$ full anti-correlation, and $\rho = 0$ no correlation. We calculated the Pearson product-moment correlation coefficient for each electrode configuration and for both BEM and FEM.

### G. Simultaneous EEG- fMRI experiment

This experiment was performed to:
1. Compare the estimated EEG current sources with the fMRI activation.
2. Compare the EEG inverse solutions between: a) the individual MRI-based model, with the exact electrode positions as identified in the MRI by the marks they leave on the rat scalp with b) the template MRI-based model with the described methodology in this paper and in Bae et al. (2015).

#### 1) Animal preparation

All procedures were performed in agreement with the policies established by the Animal Care Committee at Tohoku University (approval code, 2011AcA-40). The experiment was conducted on a male Wistar rat (Charles River, Yokohama, Japan). The rat was anesthetized with 5% of isoflurane. The scalp hair was carefully removed and the exposed skin was degreased with ethanol. The rat was orally intubated and mechanically ventilated (SAR-830/AP, CWE Inc., Ardmore, PA, USA) and a muscle relaxing agent was administered (pancuronium bromide, 2.0 mg/kg/h i.v.). The rat was placed in the prone position on the MRI-bed and the anaesthetic was switched from isoflurane to α-chloralose (80 mg/kg as an initial bolus, followed by a constant infusion of 26.7 mg/kg/h). The core body temperature was carefully regulated during the entire experiment.

#### 2) Forepaw electrical stimulation

A pair of small needle electrodes (NE-224S, Nihon Kohden, Tokyo, Japan) was inserted under the skin of the right forepaw for electrical stimulation. A block-design stimulation paradigm consisting of 10 blocks was employed, where each block comprised of a 30s forepaw stimulation followed by a 40s resting condition. A generator (SEN-3401, Nihon Kohden, Tokyo, Japan) and an isolator (SS-203J, Nihon Kohden, Tokyo, Japan) were utilized to produce the electrical pulses (3 Hz, 3 mA, and 0.3 ms width).

#### 3) EEG recordings and pre-processing

This dataset is actually previous to the 32-electrode mini-cap configuration presented in this paper. It was however acquired with a mini-cap with the sole difference of having 31 electrodes, presented by Sumiyoshi et al. (2011). The data were collected using a 32-channel MR-compatible BrainAmp system (Brain Products, Munich, Germany) with a sampling rate of 5 kHz. We applied both band-pass from 0.5 to 250 Hz and 50-Hz notch filtering. An additional channel was used to record the electrocardiogram through a needle (Model 1025, SA Instruments, Stony Brook, NY, USA) that was inserted into the right hindpaw. To synchronize the EEG data with fMRI scanning, analog outputs (5V) generated by an MRI console were digitized (16-bit resolution) as triggers to the EEG amplifiers and subsequently utilized for the offline removal of scanning artifacts. The reference and ground electrodes (NE-224S, Nihon Kohden, Tokyo, Japan) were subdermally inserted into the right and left earlobes, respectively.

The Brain Vision Analyzer software (Brain Products, Munich, Germany) was used for the offline correction of fMRI scanning artifacts. This software implements the adaptive artifact subtraction (AAS) method, in which the scanning artifact waveforms are segmented ($-10$ ms to 276 ms relative to the scanning trigger), averaged (after baseline correction), and iteratively subtracted from the EEG signals. The data were then down-sampled to 1 kHz, band-pass-filtered from 0.5 to 70 Hz using a Butterworth filter with a 24 dB/oct slope and finally exported into a binary format. The subsequent EEG data analysis was performed using custom-written software in MATLAB (MathWorks, Natick, MA, USA). To reject the ballisto-cardiogram and other undesirable artifacts from the filtered EEG data, an independent component analysis-based subtraction method was employed. The mean ERP for each experiment was computed from 900 individual somatosensory evoked potentials (SEP) (from $-50$ ms to 250 ms, time-locked to the stimulus onset).

### 4) T1 and T2-weighted MRI

The MRI data were acquired using a 7.0-T Bruker PharmaScan system (Bruker Biospin, Ettlingen, Germany) with a 38 mm diameter birdcage coil. A T1-weighted image was obtained using a 3D-RARE sequence with the following parameters: TR/ $TE_{eff}$ =300/8.5, RARE factor=4, SBW=100 kHz, flip angle=90°, FOV=34×34×30 mm³, matrix size=272×272×60 and NAX=4. T1-weighted images were utilized for the identification of the electrode positions. A T2-weighted anatomical images were obtained using the 2D-RARE sequence with the following parameters: TR/$TE_{eff}$ =4600/30, RARE factor=4, SBW=100 kHz, flip angle=90°, FOV=32×32mm, matrix=256×256, voxel size=125×125μm, number of slices=54, slice thickness=0.5mm, slice gap=0mm and NAX=24. This image was segmented and normalized using SPM8 and the template MRI (Valdés-Hernández et al., 2011). The segmentations were used to define the volumetric gray matter source space. The mid-cortex surface was calculated by shrinking the surface surrounding the brain mask, until reaching the mid layer between the pial and the white/gray interface. The skull boundaries were manually delineated. Using this head model and electrode positions, the LF matrix of the individual versions of the VolGM and SurfMid were calculated. The resulting SPM spatial normalization transformation was used to warp the fMRI activation to the template space and the atlas digitalization to the individual space.

### 5) fMRI experiment

The fMRI was acquired with the GE-EPI with parameters TR/TE =2000 ms/15 ms, SBW=250 kHz, flip angle=30°, FOV=25×14 mm² matrix=125×70, voxel size=200×200 μm², number of slices=7, slice thickness=1.5 mm, slice gap=0 mm, number of volumes=370 and dummy scans=4. The fMRI data analysis was performed using SPM5[6], which consisted of adjustments of the acquisition timing across slices, correction for head movement, and smoothing using a Gaussian kernel of 0.8 mm full width at half maximum. The single-subject analysis of the pre-processed fMRI data was performed also using SPM5 with a critical T-value for each voxel (p<0.05, FWE corrected).

### H. Further SEP experiments

We performed additional SEP experiments with forepaw, hindpaw and whisker stimulation on 6 adult Wistar rats (300-400g, Charles River, US). This time with no individual MRI. We located the electrodes on the template according to the procedure described above, based on the landmarks in the scaffold, and calculated the template LF matrices for inverse estimation.

### 1) Animal preparation

The study was approved and carried out in full appliance with Animal Care and Use Committee (IACUC) at Florida International University (IACUC 13-004). Anesthesia was induced using 5% isoflurane and maintained at 1.5-2% (mixed with 100% $O^2$ 1 L/min) during the fixation of the rat to a stereotaxic apparatus (2 SR-5R, Narishige) and the EEG mini-cap setting. The scalp hair of the rat was trimmed, shaved and rubbed with absolute ethanol (90%) for both stimulation of blood vessels and depilation of the skin. After the EEG mini-cap was secured on top of the rat head, the anesthesia was switched to dexmedetomidine hydrochloride (i.p 0.25 mg/kg, Dexdomitor, Orion Pharma). The body temperature (ML312 T-type Pod) and heart (FE136 Animal Bio Amp) and respiration (FE221 Bridge Amp) rates were continuously monitored (PL3508/P PowerLab 8/35 with LabChart Pro, AD Instruments) to guarantee a stable level of anesthesia during the entire EEG recording.

### 2) Forepaw and hindpaw stimulation

A needle electrode was inserted inside each forepaw and hindpaw for electrical stimulation. The stimulation followed a paradigm of 10 blocks with 16 s and 44 s for stimulation and resting condition, respectively. The current stimulation was delivered by the isolated pulse stimulator (Model 2100, 110V, 60 Hz, A-M Systems) with 10-ms pulse-width and 2.0-mA current amplitude at 3 Hz.

### 3) Whisker stimulation

Both sides of whiskers were used to stimulate the vibrissae system of Wistar rats. The whiskers were shortened to 1 cm, and deflections were carried out in the AP direction. Both whiskers were deflected by air puffs of 10 ms duration. The air puffs were generated from a high-pressure air tank controlled by a pneumatic picopump (PV830, World Precision Instruments) at a pressure of 14 psi. A block-design paradigm was applied. Each block consisted of 16 s of stimulation and 44 s of resting period. The air puffs were delivered at 3 Hz within each stimulation period. Each air puff was delivered using a 1 m length tube and produced an inevitable loud sound associated with the switch on/off mechanism. Visual inspection reveal not only movement on the whisker but also on the facial fur.

### 4) EEG recording and processing

A 32-channel EEG amplifier (PZ3, Tucker-Davis Technology) were used for EEG recordings. EEG signals were amplified, band filtered between 0.5-250Hz, notch filtered at 60 Hz and digitized (5kHz SR, 0.5μV resolution). Electrodes on the ears were used as ground (left) and reference (right). The data were then down-sampled to 2034.5 Hz. SEPs were created by averaging the EEG trials ($-100$ms to 300ms, time-locked to

---

the stimulus onset). These number were more than 400 for all the SEP experiments.

An in-house software (LabVIEW, AD Instruments) was used to both generate the desired triggers for the stimulation and synchronize all the recording devices (PZ3, Model 2100, and PV830) via a multiple input/output analog-to-digital converter (PCI-6259, National Instruments).

### I. EEG source imaging methods

We solved the inverse problem using the particular LF matrices of the surface model SurfMid and the gray matter volumetric model VolGM. We employed two inverse methods: 1) exact LORETA (eLORETA) (Pascual-Marqui, 2007), implemented in the fieldtrip software package[7], and the 2) LORETA-like implementation of SPM8 (SPM-LOR). For the cases described in Section I.G, we also performed a group-based inversion to increase the reliability of detecting systematic responses over subjects (Litvak y Friston, 2008).

### IV. RESULTS

For brevity, the results presented in Sections II.A and II.B correspond only to the FEM-based LF matrix. These results were not done for the BEM-based LF matrix, since the goal of this paper is to probe the validity of rat ESI and not the possible differences between both methods. However, it is has been demonstrated that FEM and BEM forward models with only three conductivity compartments are very similar (Vorwerk et al., 2012).

### A. Electrode resolution

As we have suggested, among all lead centers, there is a repertoire of different behaviors in the HSV and FSV curves versus lead separation which we shall not describe for brevity. In general, the main difference between these curves and those for spherical models is that the former decrease after a certain lead separation, while the latter saturate. This can be seen in Figure 2 for the ML and AP directions. A detailed discussion about the reasons for this is beyond the scope of this paper.

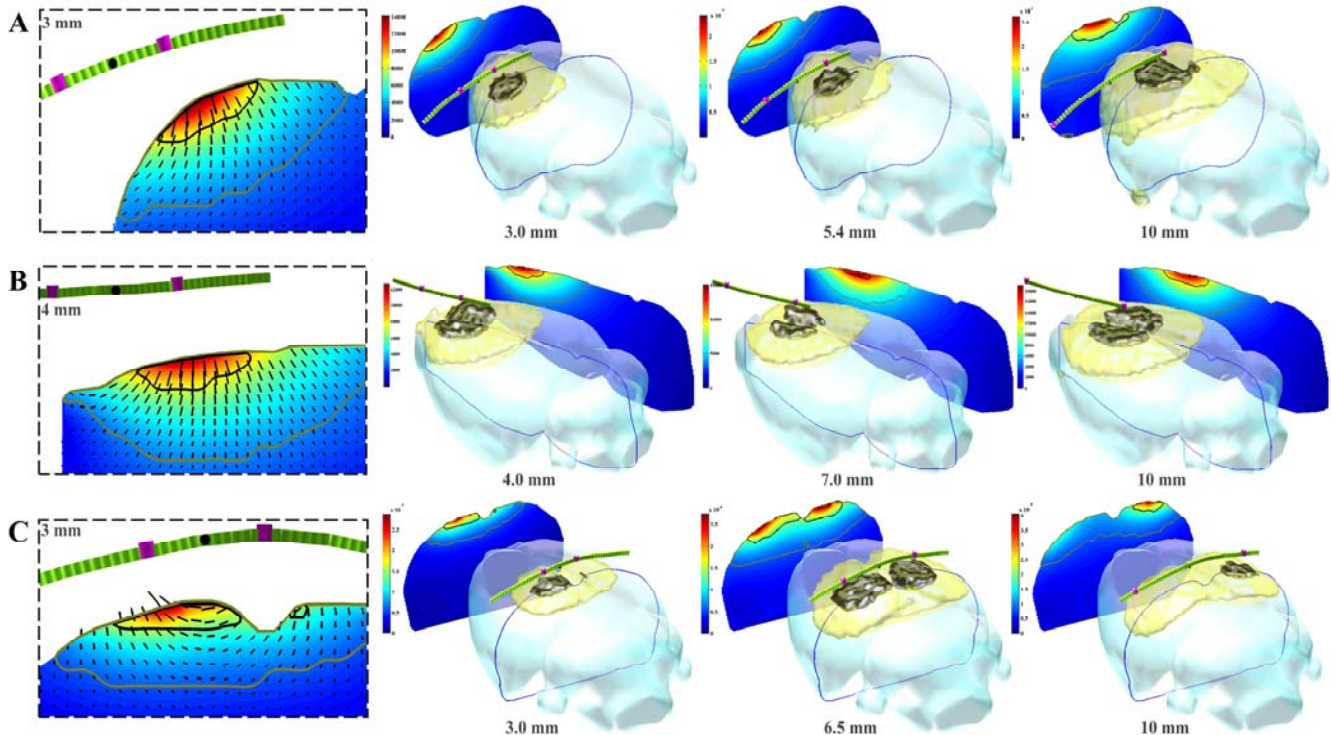

Fig. 3. **Sensitivity behaviour for increasing separating leads (green cylinders) for a ML center (black dot).** The results for four lead separations are shown, corresponding to the leads represented with the magenta cylinders. To give an idea of the extension and depth of the HSV and FSV regions relative to the brain volume we represent them with dark and light brown surfaces respectively, while the brain volume is represented with a light cyan surface. For each separation case, the scalar sensitivity (values according to the colorbar) of the slice containing the leads was projected outside the brain to ease the illustration. The intersections of this slice with the HSV, FSV regions and the brain are represented with a black, light brown and blue curves respectively. To the left, a zoom of the sensitivity profiles for 2 mm and 15 mm are shown. For the former the lead is more sensitive to tangential dipoles while for the latter the lead is completely sensitive to radial dipoles.

---

[7] online at http://www.fieldtriptoolbox.org/

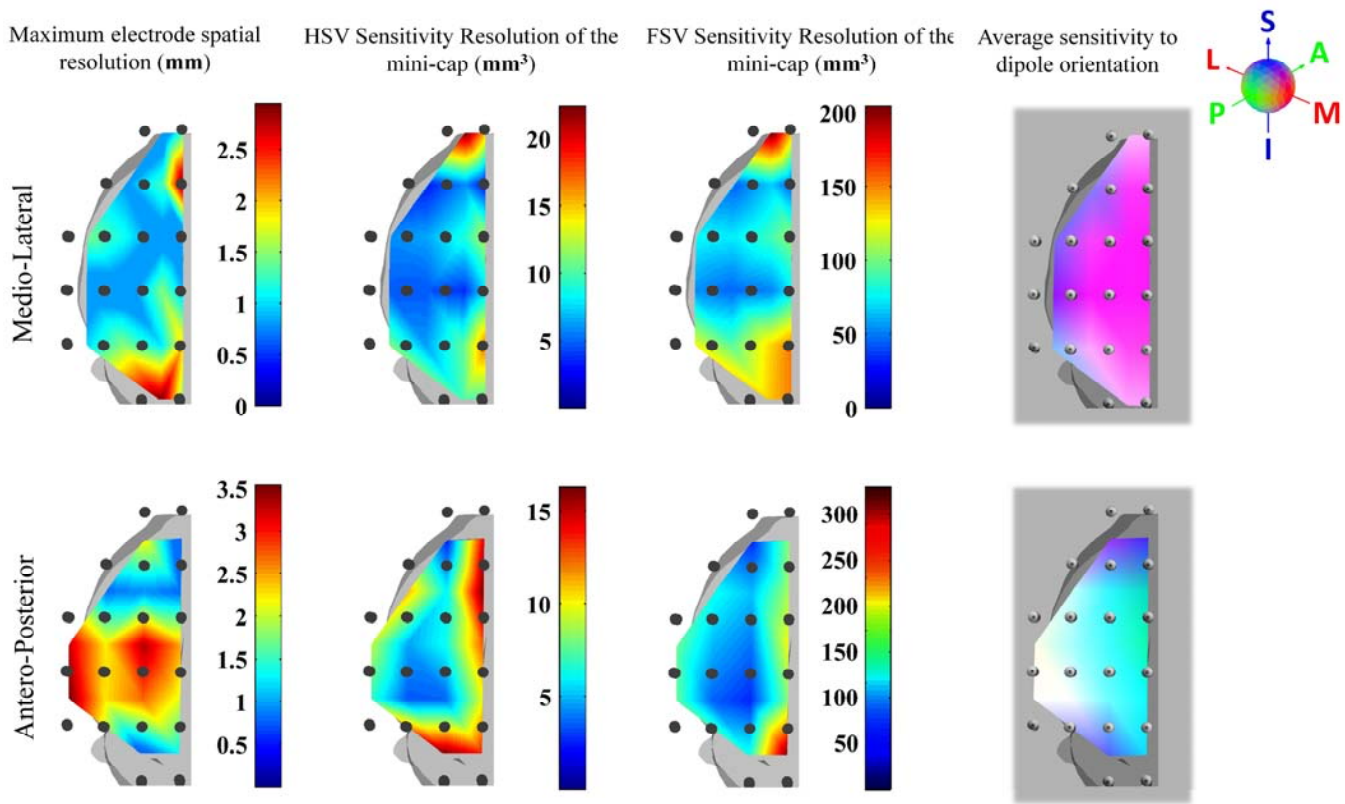

`Fig. 4. **Results of the HSV/FSV-based electrode resolution for scalp EEG in rats**. The values of the topographies at each location indicates the interpolated result for a pair of electrode at that location. The results are shown only for one hemisphere for obvious symmetry reasons. The first column shows the topographies of the maximum electrode spatial resolution, i.e. the minimum electrode separation for which a pair of electrode is not measuring redundant information. This value is the maximum, i.e. the worst, between those obtained from the HSV and FSV curves. They are also bounded from below by the electrode size of our mini-cap (0.8 mm). Since the separations between AP and ML electrodes pairs of our EEG mini-cap are 3 mm and 4 mm respectively, the cap is not measuring redundant information. The second and third columns report the sensitivity resolution of our rat EEG mini-cap. The fourth column shows the main orientation of the sensitivity. This is represented with the RGB code. Red: ML, green: AP and blue: IS. As expected the ML leads in the middle of the measurement region are mainly sensible to tangential ML oriented dipoles, while in the boundaries to radial dipoles. The central AP leads are more sensitive to tangential AP dipoles, while anterior and posterior AP leads are more sensible to radial dipoles.

The only cases worthy of discussion are those centers near the lateral, the anterior and the posterior boundaries of the source space. In those cases, the decrease, especially of the HSV, starts early in the curve even at lead separations below those of the electrode pairs of the EEG mini-cap. This is because the contribution to the HSV/FSV is mainly from only the electrode closer to the source space, for the other is already too far. This has two implications: 1) the sensitivity resolution, as quantified by the HSV/FSV results better than expected and 2) the sensitivity profile is more radial, or more properly said since we are not dealing with spheres, oriented toward the electrodes. Note that the AP example depicted in Figure 2 is one of these cases: the lead center is one of the most posterior and the HSV/FSV curves decrease after 4 mm.

Figure 3 illustrates the behavior of the sensitivity for three lead centers. We chose a ML center close to the lateral boundary of the mini-cap (Figure 3A) and an anterior AP center (Figure 3B) to demonstrate how they are affected by the aforementioned electrode-source separation effect. For these cases, the sensitivity profile is completely due to one of the electrodes and thus radial at the corresponding separation of the mini-cap. There is another example in Figure 3C corresponding to a ML center close to the medial plane and central. For this case, the sensitivity profile at the corresponding electrode separation of the mini-cap is still tangential with radial contributions. Figure 3 is also intended to illustrate how the HSV, FSV and sensitivity evolves with lead separation.

We consider that, even in the presence of these realistic rat geometry effects, the criteria to assess the maximum electrode spatial resolution (Ferree et al., 2001) can remain the same, for differences in the HSV/FSV still mean non-redundant information. We used both HSV and FSV since they provide some complementary information. The former gives a better measure of which source space region is mostly contributing to the sensitivity. The latter is less sensitive to geometry and provides a better measure regarding how much of the whole brain is actually measured by the electrodes. This can be seen in Figure 3.

Figure 4 summarizes the results for all electrode pairs and

This is consistent with the LFs observed in Figures 3A and 3B.

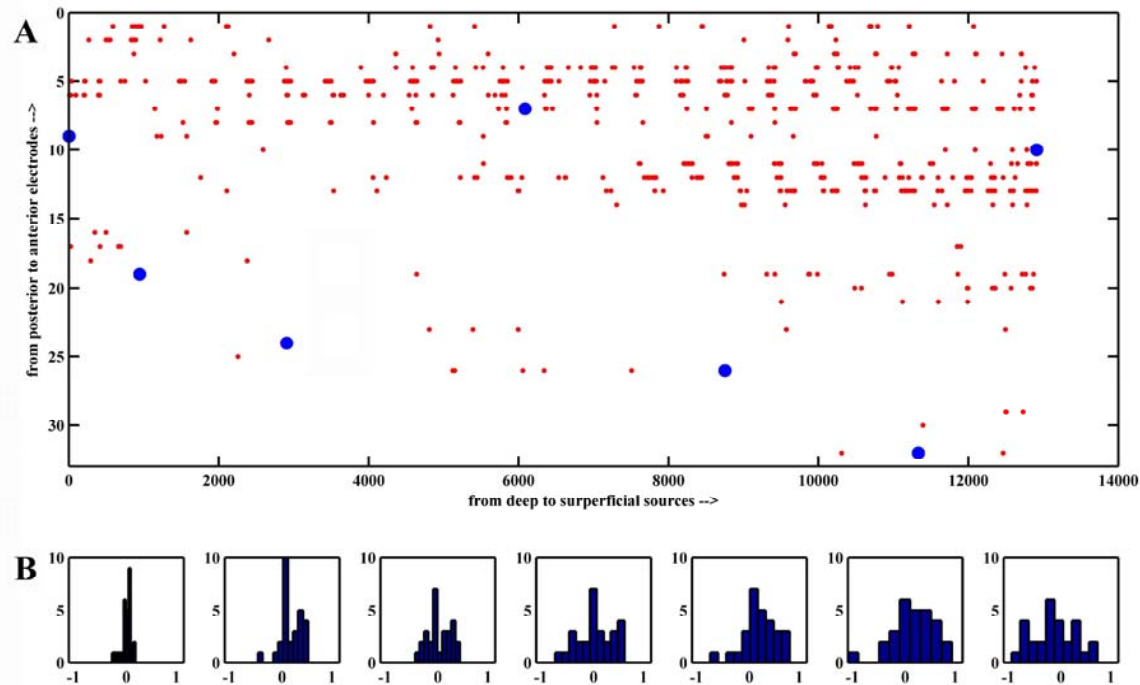

Fig. 5. **Demonstration of the Gaussian behaviour of the elements of the LF matrices across individuals and electrode configurations**. A) Matrix of the same size of the LF matrix with the result of the Smirnov-Kolmogorov test under the null hypothesis of Gaussianity (p<0.05). The red dots (which were augmented for visualization purposes) identify the elements where the null hypothesis was rejected. We note that the Gaussianity is violated more toward deep sources and anterior electrodes. This might imply an overestimation of the CRLB for deep and anterior dipoles. However non-Gaussian elements are barely the insignificant 0.17% of the total number of elements. B) Histograms of randomly chosen elements which are identified with blue dots in the matrix above. The histograms follow the same order, from left to right, of their corresponding blue dots.

corresponding lead centers. Assuming smoothness of the results, we interpolated the values over the whole measurement region to create scalp topographies for better visual understanding. The first conclusion is that all electrode pairs in our rat EEG mini-cap are measuring non-redundant information. A useful information for cap builders is that the limit for cap density in rat scalp EEG is lower in the boundaries of the region, and probably is worse beyond. The electrode resolution, as quantified by the HSV and FSV values, is worse in the boundaries of the mini-cap as well. Although the maximum HSV is higher for the ML topography than for the AP topography, the HSV and FSV resolutions along the AP direction are in general worse than along the ML direction.

The topographies of the average sensitivity orientation, as calculated by Eq. (5), are also shown in Figure 4. We used the RGB representation for this vector and the intensity of the color represents the LF coherence as quantified between 0 and 1. We note that for electrode pairs far from the boundaries, the main orientation of the sensitivity is a mixture of tangential and radial contributions. This corresponds to magenta (red+blue) and cyan (green+blue) colors for ML and AP orientations, respectively. This is consistent with the LF observed in the medial ML center depicted in Figure 3C. On the other hand, the main sensitivity orientation for electrode pairs closer to the boundaries is more along the IS direction (bluer color), i.e. they are more radial.

Note that, for the AP centers in the lateral boundary, the main orientation is a combination of the three ML, AP, and IS directions (whiter color). We point out that this does not mean there is a mixture of sensitivity orientations resulting in an isotropic sensitivity profile. It means that the main orientation is in one of the octants of the 3D space. This is the average of a current field that flows from the source electrode along both the ML and IS directions down to the source space, taking the AP direction and returning again along both the ML and IS directions up to the sink electrode. The intensity of all vectors is above 0.7, denoting the expected high coherency in the LF and confirming that we are indeed mapping a coherent "main field direction" within the HSV.

### B. Mapping the Cramér-Rao Lower Bound

Figure 5 demonstrates the zero-mean Gaussian behavior of the elements of the LF matrix across individuals and electrode configurations. Only the 0.17% of the elements are non-Gaussian. Therefore, we consider the zero-mean normality assumption justified.

To report RMS and Z we chose a value for $f \equiv \sigma/Q$. We used an estimate of $\sigma$ obtained from the SEP experiments presented in this paper. After averaging across trials, the absolute value of the signal across the pre-stimulus interval and across experiment is about $2.5 \pm 1.0 \mu V$. On the other hand, the equivalent current dipole in whisker stimulation was measured in the barrel cortex (Riera et al., 2012), being $Q \sim 10 nAm$ at around $50 ms$, where the peak occurs. This yields $f \equiv \sigma/Q \approx 250\Omega/m$.

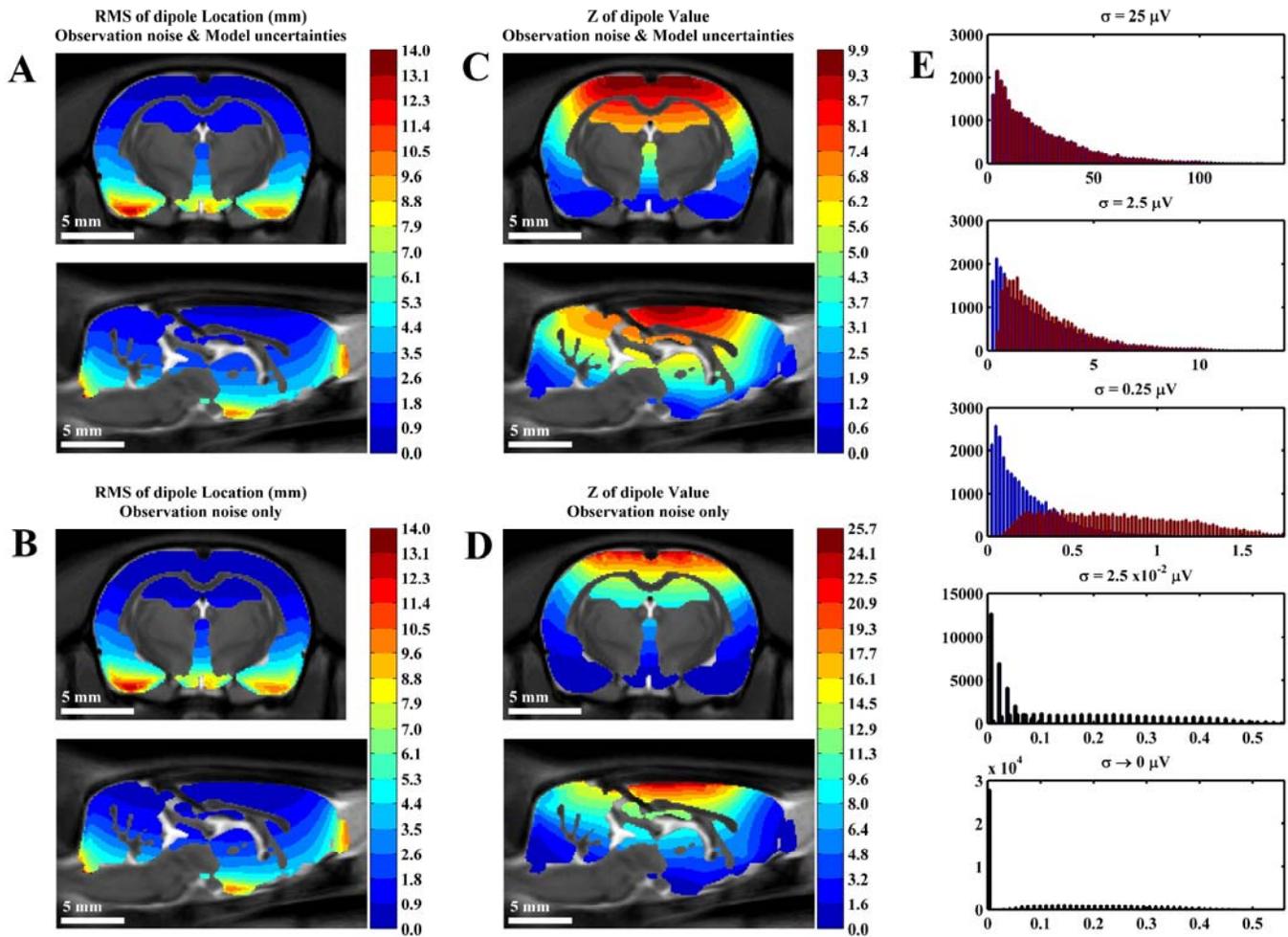

Fig. 6. **CRLB results for the single dipole case**. A, B, C and D corresponds to an observational noise level of $\sigma = 2.5\mu V$. A) RMS for the forward model considering both noise and model uncertainties. B) RMS of the noise only case. C) Z for the noise and model uncertainty case. D) Z for the noise only case. Each value in the image corresponds to the results for a dipole located at that point. Only the points in the gray matter source space (VolGM) model are shown since this is usually the only space where dipoles are fitted. The medial sagittal and mid-axial slices are shown. Note that the RMS between A and B look similar, while for Z in D is bigger than in C. E) Histograms of the RMS for the noise only case (blue bars) and the noise and model uncertainty case (red bars) for various values of $\sigma$ (growing from top to bottom). Note that for low the RMS is completely due to model uncertainties. For $\sigma = 2.5\mu V$ the RMS distribution is slightly higher for the model uncertainty case. For high $\sigma$, both cases are completely dominated by noise, being identical.

In the following figures, we summarize the values of RMS and Z by only reporting their average across X, Y and Z dipole orientations. Figure 6 shows the values of RMS and Z for a single dipole located at each point of the source space model of VolGM. As expected from the high co-planarity in the electrode configuration, and due to the dependency of the values with the dipole-electrode distance, the values of RMS and Z approximately vary in layers from superficial to deep sources, with the latter being very difficult to estimate correctly.

In the range of the noise of ERP experiments, the use of the template as a surrogate of the individual implies a slight increase in the lower bound of localization errors with respect to the use of a model only affected by noise. The point-wise difference, averaged across all points, is $0.48 \pm 0.22 mm$. With the decrease of $\sigma$, the RMS for the model uncertainty case reaches a limit around 0.05 mm, which is far below the best localization error that any known inverse method would provide.

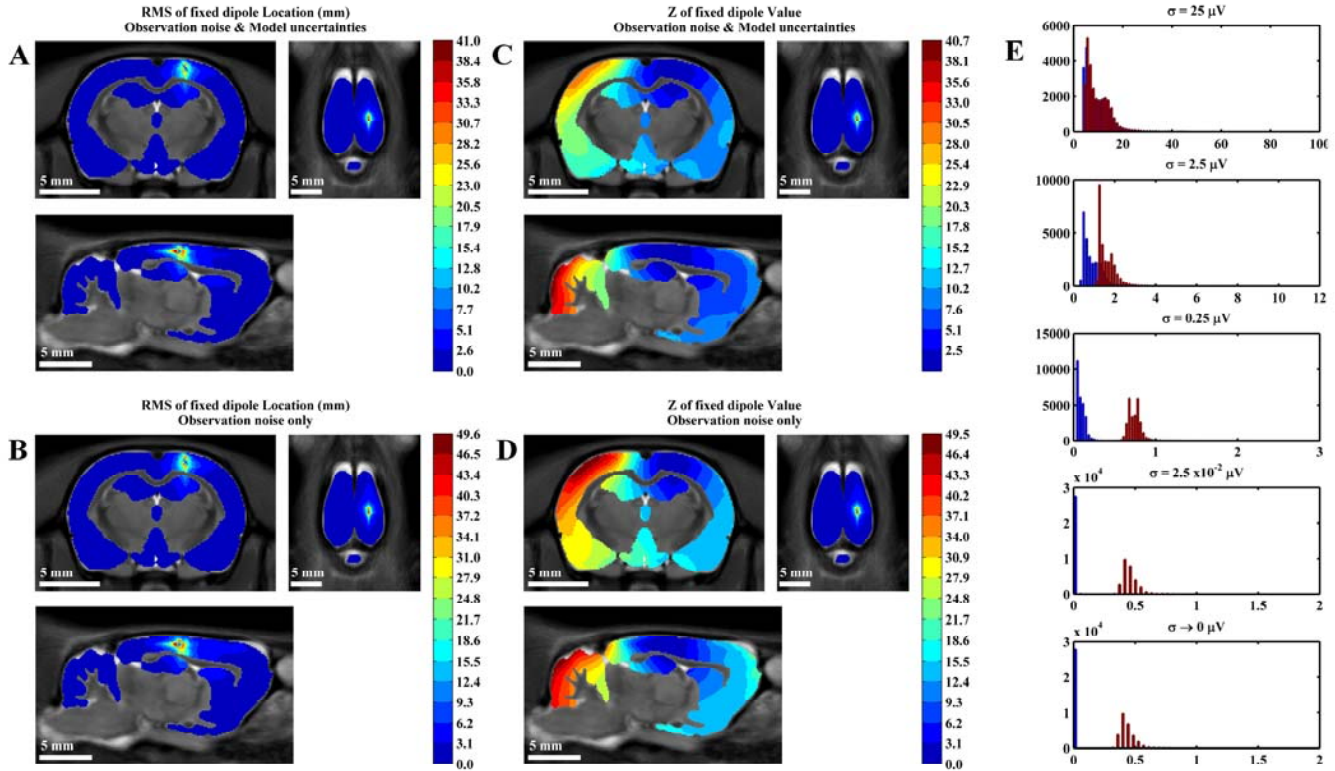

**Fig. 7. CRLB results for the double dipole case.** One dipole is fixed and oriented along Z, and the second is moving across all points of the source space. The results shown are those of the fixed dipole. A, B, C and D corresponds to an observational noise level of $\sigma = 2.5\,\mu V$. A) RMS for the forward model considering both observation noise and model uncertainties. B) RMS of the noise only case. C) Z for the noise and model uncertainty case. D) Z for the noise only case. The value at each point of the image corresponds to the results for a fixed dipole when the moving dipole is located at that point. The brain slices are the ones containing the fixed dipole. E) Histograms of the RMS for the noise only case (blue bars) and the noise and model uncertainty case (red bars) for various values of $\sigma$ (growing from top to bottom). Note that for low $\sigma$ the RMS is completely due to model uncertainties and cannot be lower than about 0.3 mm. For high $\sigma$, both cases are completely dominated by noise, being identical. We note that the RMS for model uncertainties (A) is in general higher than for the noise only case (B), as evidenced in the histogram corresponding to $\sigma = 2.5\,\mu V$.

With the aid of our digitalization of the Paxinos and Watson atlas, we summarized in Table 2 the mean and standard deviation of RMS and Z across the points belonging to each rat cortical region, for $\sigma = 2.5\,\mu V$. For brevity, among the 96 structures of the atlas, we only present some of those related to the somatosensory evoked potential (SEP) experiments of this paper. The results for the rest of the cortex is summarized using conglomerates of structures with functional relation. These were also formed attending to closeness and similar depth, based on the fact that depth is probably the most determinant variable on the results. For example, the secondary somatosensory system was left apart from the whole somatosensory system because the former is significantly deeper than the latter. Likewise, we separated the parietal association cortex from the parietal system. On the contrary, other systems like the visual or auditory systems were formed by grouping their corresponding primary and secondary structures since they are all at approximately the same depth.

In order to quantify how large is the RMS compared to the size of the anatomical structure containing the dipole, we defined the Relative Linear Error (RLE). This was calculated for each anatomical structure of the atlas as:

$$RLE = \frac{RMS}{\sqrt{L_{AP}^2 + L_{ML}^2 + L_{IS}^2}} \times 100\% \ , \quad (14)$$

where the denominator is the root-mean-square of linear size of the structure, taken across the AP, ML and IS dimensions. Each linear size is the length of the structure along that dimension. We also report the RLE in Table 2. For SEP experiments, if the dipole would be located at the expected somatosensory region, its RMS would be roughly between 10% and 20% of the root-mean-square of the linear size of the cortical region.

Figure 7 shows the values RMS and Z in the double dipole case for the fixed dipole. This is oriented along the Z direction. For brevity, we chose only an example where the fixed dipole is located in a superficial point of the cortex, with orientation along Z. The results from placing a fixed dipole at a deep point can be envisaged by inverting the view and analysing the current results when the moving dipole is deep. For brevity, we omit the analysis resulting from placing both dipoles in deep regions, but it is highly likely that they cannot be estimated.

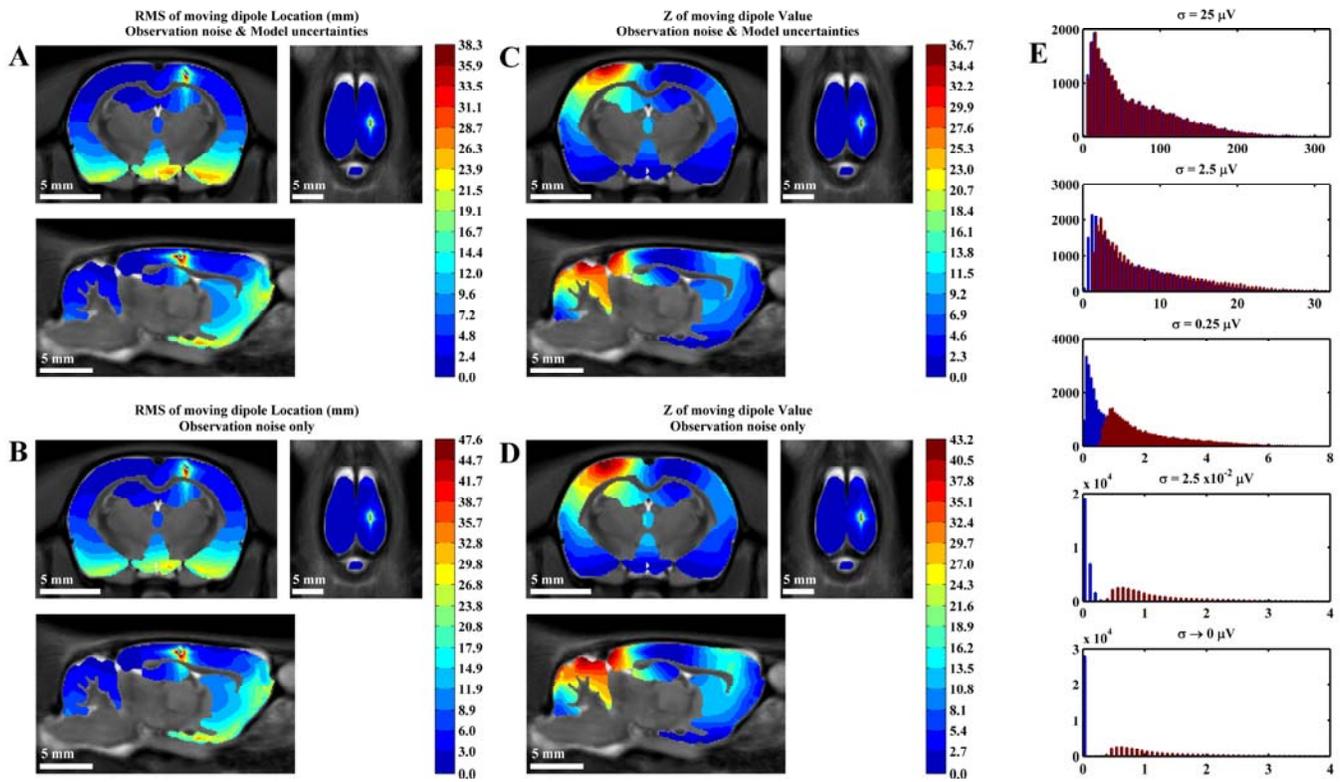

Fig. 8. **CRLB results for the double dipole case.** In this situation we simulate two dipoles. One is fixed and the second is moving across all points of the source space. The results shown are those of the moving dipole. The results shown are those of the fixed dipole. A, B, C and D corresponds to an observational noise level of $\sigma = 2.5\mu V$. A) RMS for the forward model considering both observation noise and model uncertainties. B) RMS of the noise only case. C) Z for the noise and model uncertainty case. D) Z for the noise only case. Each value in the image corresponds to the results for a moving dipole when it is located at that point. The slices are the ones containing the fixed dipole. E) Histograms of the RMS for the noise only case (blue bars) and the noise and model uncertainty case (red bars) for various values of $\sigma$ (growing from top to bottom). Note that for low $\sigma$ the RMS is completely due to model uncertainties and cannot be lower than 0.3 mm. For $\sigma = 2.5\mu V$ the RMS distribution is slightly higher for the model uncertainty case. For high $\sigma$, both cases are completely dominated by noise, being identical.

For the typical experimental noise level ($\sigma \approx 2.5\mu V$), the point-wise RMS difference between the "noise only" or "model uncertainties" cases, averaged across positions, is $0.67 \pm 0.23mm$. The average separation between dipoles for which $Z \leq 1$ are $2.38 \pm 0.99mm$ and $3.14 \pm 1.45mm$ for the only noise and model uncertainties cases, respectively. This means that, on average, below these separations the fixed dipole becomes undetectable by any estimation method. Besides, using the template as a surrogate of the individual model implies a fixed bias in the RMS around $0.3mm$ which will not improve with lower $\sigma$.

Figure 8 shows the results for the moving dipole. For the typical experimental noise level ($\sigma \approx 2.5\mu V$) the point-wise RMS difference between the noise only and model uncertainties cases, averaged across all positions, is $1.40 \pm 0.89mm$. In both cases, the distribution of dipole separation (histogram not shown) for which $Z \leq 1$ are clearly bimodal. This corresponds to two spatial clusters of the position of the moving dipole. The first spatial cluster is compact, approximately isotropic and it is localized around the fixed dipole. It corresponds to the positions of the moving dipole that are "too close" to the fixed dipole. The second is widespread and it is distributed around the deepest source locations. It corresponds to the positions of the moving dipole that are "too deep". Both clusters can be seen in the sagittal and coronal slices of Figure 8 (bluest colors). The average spatial extension of the first cluster is $2.52 \pm 1.06mm$ and $2.88 \pm 1.11mm$ for the only noise and model uncertainties cases, respectively. Below these separations the moving dipole becomes undetectable by any estimation method. Besides, using the template as a surrogate of the individual model implies a fixed bias in the RMS around 0.3-2 mm, depending on the location of the moving dipole, which will not improve with lower $\sigma$.

*In summary,* double dipoles cannot be estimated for average separations below $3.14 \pm 1.45mm$ and using the template as a surrogate of the individual model implies an extra localization error of about 0.5 mm that cannot be reduced by decreasing the $\sigma$.

### C. Comparisons of the LF matrices

Figure 9 shows the comparison between the individual LF matrices and the template LF matrix. To quantify how close the Pearson product-moment correlation coefficients are to 1, their cumulative distribution function across source points, electrode configuration and individuals are characterized with their 5 and 95 percentiles. These are $\tilde{\rho}_i = 0.97$ and $\tilde{\rho}_1 = 0.99$ for BEM; and $\tilde{\rho}_i = 0.97$ and $\tilde{\rho}_1 = 0.99$ for FEM. These are very high coefficients, as it can be visually checked in the corresponding correlation plots in Figure 9. These values provide evidence for the use of either BEM or FEM template LF matrices for ESI in rats. These high correlations are actually not surprising since rat heads are very similar.

### D. Simultaneous EEG-fMRI experiment

Figure 10 shows where the fMRI activation occurs. The SPM T-map for values above 3 is shown. The activation mostly falls within S1FL, contralateral to the stimulated paw, though a small part is shared with the central region of M1. We shall refer to the locations of the brain as contralateral and ipsilateral to the stimulated side.

The SEP resulting from averaging the 900 trials is shown in Figure 11. We chose six time instants to show their respective topographies. These instants were chosen after inspecting the forthcoming EEG inverse solutions. We consider them as the instants when the most relevant spatiotemporal features of the solution occur. No wonder that, excepting instant $55ms$, they coincide with the peaks of the SEP. In fact, some of these peaks are well described and reproducible components of SEP in paws stimulations. The first is P1. It is positive and occurs at 15ms. The second, N1, is negative and occurs at 25ms. The third is P2 and occurs at 65ms. They have well localized contralateral topographies. These peaks have been identified for electrical forepaw stimulation elsewhere, e.g. Franceschini et al. (2008). Like in that paper, were could not identify the contralateral N2 component.

On the other hand, there is a marked ipsilateral positivity at 25ms, though its amplitude is lower and its topography is less localized than that of N1. Moreover, there is a localized ipsilateral negative component at $45ms$. It seems as if there would be a translation of the N1 component toward the ipsilateral hemisphere during the period between $25-45ms$. This is suggested by the topography at $35ms$. The inverse solutions for these instants are shown in Figures 12 and 13.

By significant *ESI activity*, we considered the absolute value of the solution which is above the 99 percentile at each instant and above 25% of the overall maximum across spatial points and instants. The latter restriction is based on the CRLB results. Table 2 shows that the range of $Z$ for somatosensory regions is ~6-8. However these values were calculated assuming the observed maximum dipole amplitude Q in somatosensory responses. Based on the rough relation $Z \propto f^{-1} \propto Q$, equivalent dipoles with amplitudes 6-8 times smaller than this maximum, cannot be correctly estimated. This does not preclude the estimation of spurious inverse solutions. Based on this rationale, but being cautious, we decided to discard the inverse solutions with an absolute value 4 times (25%) smaller than the overall maximum. Finally, the overall minimum of the thresholded absolute value was subtracted and the resulting values normalized to 1, by dividing by the overall maximum.

The ESI activity for the SurfMid model are shown in Figure 12. This figure contains the results for the template and individual models and for the eLORETA and SPM-LOR methods. For both inverse methods, the ESI activity appears around contralateral S1FL at $14ms$ and peaks at $15ms$, disappearing after 16 ms. At 20ms the ESI activity reappears around contralateral S1FL, with the maximum peak at 25 ms, and lasts until 36 ms. Ipsilateral ESI activity appears at 41 ms in the anterior region of contralateral M1 and in the region occupied by the contralateral secondary somatosensory area which moves toward S1FL until the end of the analyzed interval (5 ms-70 ms).

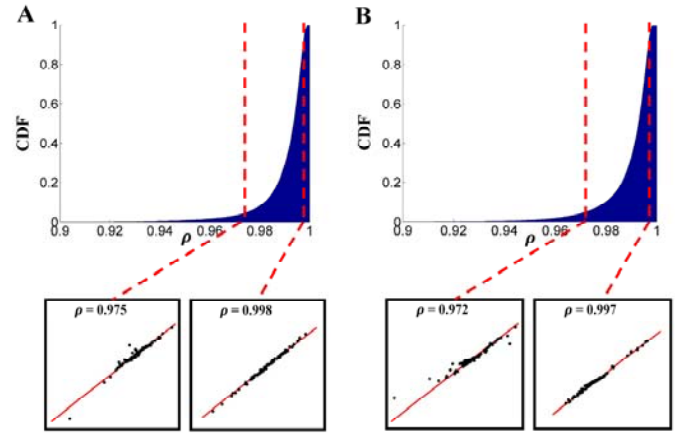

Fig. 9. **Pearson product-moment correlation coefficient between individuals and template LF matrices**. Column A and B show the results for the BEM and FEM respectively. Top row: cumulative distribution function (CDF) across individuals, electrode configurations and source points. Bottom row: plots of the values of the individual LF matrix vs. template LF matrix at the 5 and 95 percentiles of the CDFs.

The main difference between both inverse methods is that while eLORETA localizes the ESI activity almost completely within and centered in the S1FL region, SPM-LOR shifts the center toward the central part of M1. The second main difference between both methods is that the ipsilateral ESI activity at 45 ms is more extended and anterior for SPM-LOR than for eLORETA.

Figure 13 shows the solution for the volumetric model (VolGM, see Table 1) using SPM-LOR. The results essentially resemble those obtained using SurfMid in Figure 12. We do not show the eLORETA solution for the volumetric model because we consider the solution was plainly non-sense, for the ESI activity appeared at too deep regions and spread into exaggerated volumes toward the olfactory bulb. This might be a consequence of this type of standardized LORETA methods, which only guarantee zero localization error of the maximum of the solution, ignoring the whole spatial distribution (Pascual-Marqui, 2007).

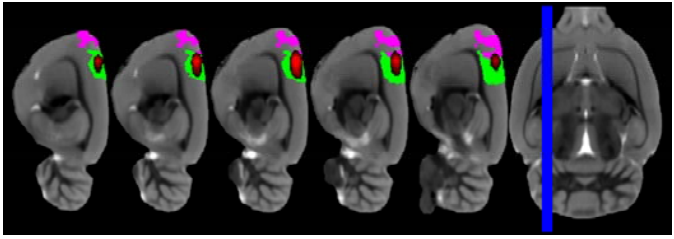

Fig. 11. **Sagittal slices depicting the fMRI activation after the forepaw stimulation in the EEG-fMRI experiment**. Red: fMRI activation. Green: S1FL. Magenta: M1. Both the fMRI activation area and atlas structure are contralateral to the stimulated forepaw. It can be seen that the activation is mostly within S1FL but a small part appears in M1.

There is a great similarity between the ESI activity estimated using the template and the individual LF matrices in Figures 12 and 13. We remind the reader that the individual solutions were estimated using the LF matrix calculated from the head model extracted from the individual MRI; after accurately identifying the electrodes in the MRI by the marks they leave on the rat scalp. On the other hand, the template solution was calculated using the template LF matrix calculated with the strategy described in this paper and in Bae et al. (2015).

The polarity of the SEP components P1, N1 and P2 is mirrored by the corresponding ESI activity using both methods. For the surface model, the ESI activity is negative for 15ms and positive for 25*ms* and 65*ms*. For the volumetric model, the estimated distributed dipoles corresponding to the ESI activity are oriented inward for 15ms and outward for 25*ms* and 65*ms*, almost all perpendicular to the cortical layers.

*E. Further SEP experiments*

Due to electrolytic shortcut provoked by the smearing of the conductive gel, we discarded two rats from this dataset. Figure 14 shows the average across rats and hemispheres of the EEG inverse solutions after the forepaw, hindpaw and whisker. For obtaining these group-based inverse solutions, we used the method described in Litvak and Friston (2008) and implemented in SPM8. For brevity, we shall only depict the systematic effects in the sample, and not the individual solutions. However the latter very much resemble their averages. We have also restricted the presentation of the results for the volumetric model since both volumetric and surface solutions using SPM-LOR are alike, as we could verify in the EEG-fMRI experiment. In this respect we also deem the volumetric solution as more descriptive of the whole process, allowing for the estimation of both unconstrained dipole orientations and deep sources.

Similar to the EEG-fMRI experiment, the forepaw ESI activity appears at 14 ms in both contralateral S1FL and in a small portion in the central M1. The ESI activity reappears later and peaks at 26 ms, centered at the S1FL. Then it attenuates, spreads and splits toward anterior and posterior regions. The ESI activity resurges during 50-70ms. These patterns resemble those obtained from the EEG-fMRI experiments. For the hindpaw stimulation, the ESI activity also appears at 14ms within the contralateral S1HL and remains concentrated in this region until it splits toward M1 after 23ms. The maximum peak is reached at 33ms mainly at both the contralateral S1HL and M1. The ESI activity continues and spread until 55ms. For the whisker stimulation, the first activation appears at 15ms in the contralateral Pt. It continues and spreads toward the contralateral S1BF, peaking at 48ms. After this instant the ESI activity is shared by both structures until the end of the interval, but mainly within Pt.

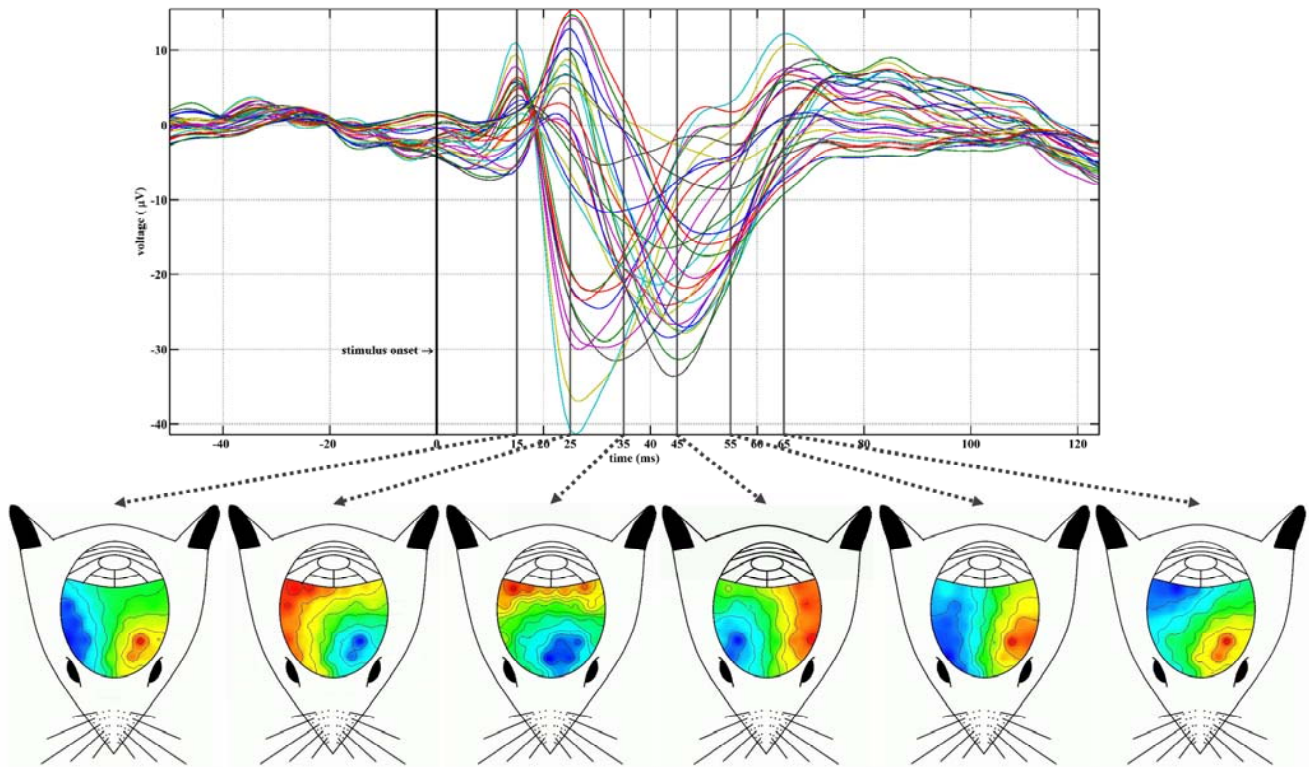

Fig. 10. **Elicited SEP (32 channels) after forepaw stimulation in the EEG-fMRI Experiment**. This is the average of 900 trials. Time is locked to the stimulus onset. The topography (in a schematic model of the actual measured region on the rat scalp) of six time instants are shown. These are the same instants chosen for depicting the inverse solutions in the next two figures. As can be seen in the SEP, these time instants were not chosen arbitrarily but coincide with the main maxima and minima of the signals. Three components of the SEP can be identified: P1 (15ms), N1 (25ms) and P2 (65ms).

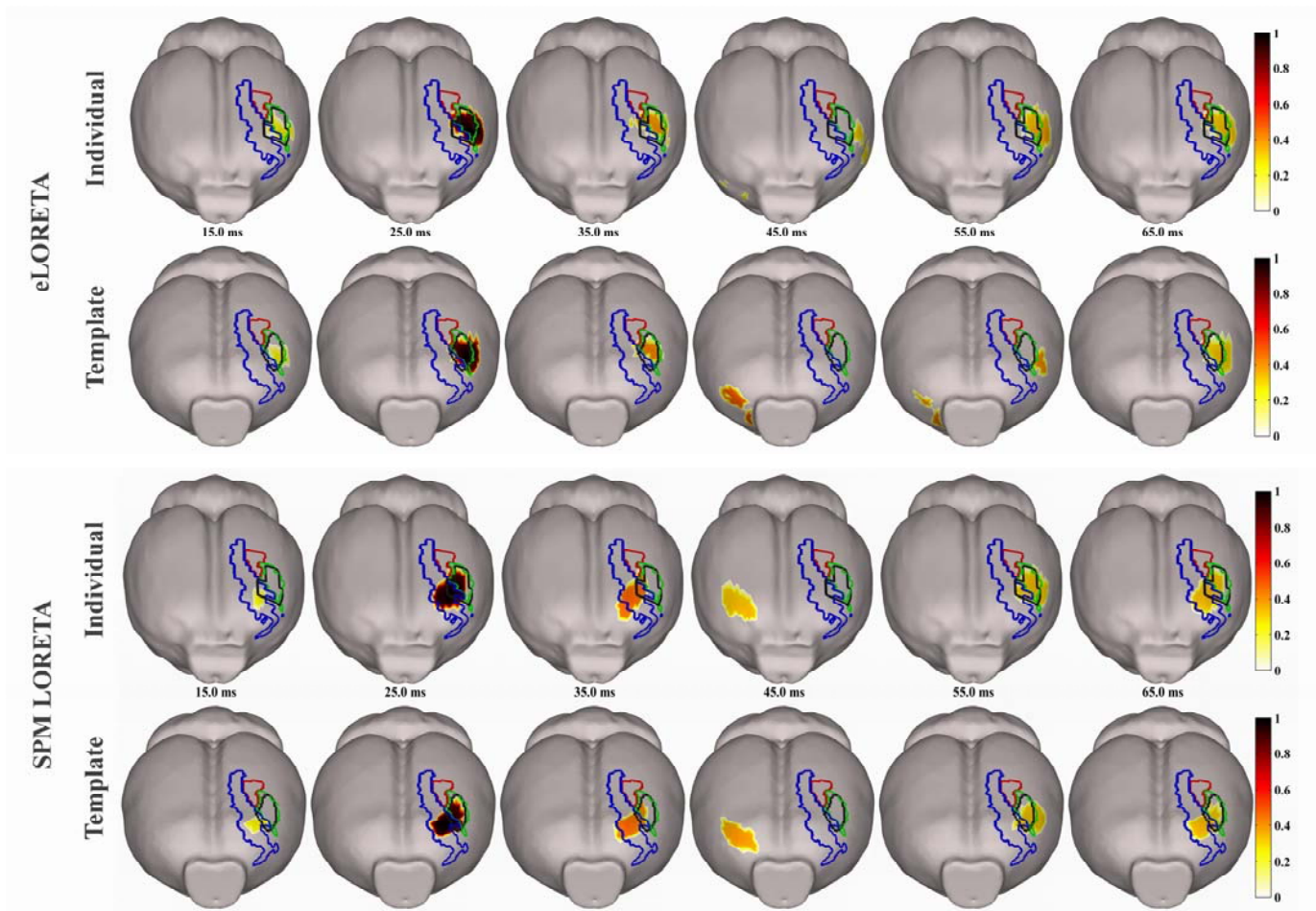

Fig. 12. **ESI in the mid-cortex surface model for the forepaw stimulation in the EEG-fMRI experiment**. The eLORETA and SPM-LOR methods are presented. The absolute value of the inverse solutions are shown for both the exact individual model (with exact MRI-based electrode positioning) and the template model built as described in this paper. The green, red and blue contours delineate the intersection of this surface with the volumes defined by S1FL, S1HL and M1. The black contour is the intersection of the surface model with the volume formed by the T-map activation for values above 3. Only the 99 percentile of the values of the inverse solutions are shown. Besides, values below 25% of the total maxima across surface vertices and time points were also discarded to avoid spurious inverse solutions at instants of no expected ESI activity. The color bars linearly map the values from the lowest to the highest value across all vertices and time instants. Evidence for the use of the template as a surrogate of the individual is shed by the great similarity between the solutions using the individual and the template models. If we expect that the solutions for the N1 and P2 components of the SEP coincide with the fMRI activation, it seems that the eLORETA outperforms the SPM-LOR. The solutions look in some agreement with what is described in the literature using VSD.

## V. DISCUSSION

### A. Interpretation of the results. From human to rat scales.

It is the usual practice in rodent Neuroimaging to multiply the dimensions of the voxels by 10, to be dealt with by the usual programs and algorithms used in human Neuroimaging, such as tissue segmentation, spatial normalization and fMRI analysis (Ashburner y Friston, 2005, 2000), among others. This is justified by a rough scaling of one order of magnitude between human and rat head sizes. Indeed the linear sizes can be easily obtained from an MRI. For the adult human brain, they are about 13 cm, 13 cm and 18 cm for the ML, IS and AP directions, respectively. For an adult rat brain they are about 16 mm, 11 mm and 24 mm, respectively.

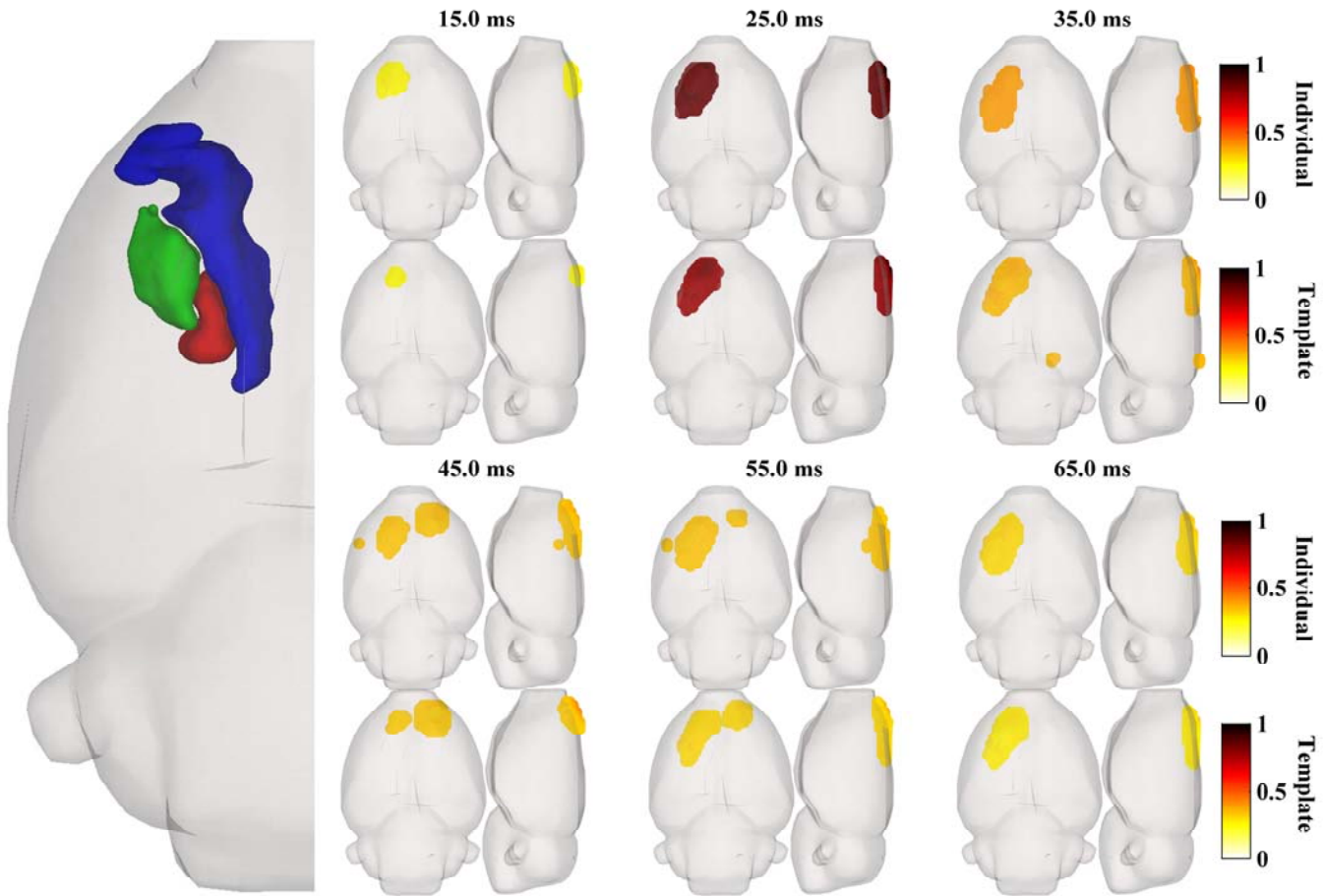

Fig. 13. **ESI in the volumetric gray matter volume for the forepaw stimulation in the EEG-fMRI experiment using the SPM-LOR.** The absolute value of the inverse solutions are depicted for both the exact individual model (with exact MRI-based electrode positioning) and the template model built as described in this paper. The light gray surface represents the brain. To the left, the green, red and blue surfaces represent S1FL, S1HL and M1, respectively. Only the 99 percentile of the values of the inverse solutions are shown. Besides, values below 25% of the total maxima across surface vertices and time points were also discarded to avoid spurious inverse solutions at instants of no expected ESI activity. The color bars linearly map the values from the lowest to the highest value across all vertices and time instants. The results very much resemble those for the surface model with this inverse method.

Downscaling an ESI head model by 10 only implies multiplying the LF matrix by $10^{-2}$. Thus, the relative spatial distribution of electrode sensitivity, and the resulting HSV and FSV do not change. Besides, the RMS is downscaled by 10 and the Z remains unchanged, yielding exactly the same conclusion regarding EEG and ESI.

However, the rat head is not a plain 10-downscaled version of the human head. In fact, the analysis done for spheres fitted to human heads does not apply for rats, for the rat head and its conductivity compartments are very unlike to those of the spherical model, as can be seen in Figure 1B. Besides, the area covered by the rat EEG mini-cap is more reduced and less curved than that of standard human head EEG montages. Furthermore, the relative skin/skull thickness is not the same between adult humans and rats, and does not distribute similarly and evenly across the head. These are the reasons that led us in this paper a) to evaluate the electrode resolution for our rat EEG mini-cap, b) to investigate the sensitivity profile of the electrodes of the mini-cap, and c) to calculate the lower bounds for the source estimation errors under the template-based forward models in Eqs. (1) and (B1); all this after realistically modeling the rat head with the three conductivity compartments of brain, skull and skin.

Nevertheless, the 10-scale rule can be used to approximately validate the range of the results of this paper. The first check is with the values of the LF matrix. The maximum value for the rat is $\sim 33000\,\Omega/m$ while for the sphere model fitted to human head (brain sphere $8\,cm$) it is about two orders of magnitude smaller, i.e. $\sim 460\,\Omega/m$.

An important question is whether our rat EEG mini-cap is measuring signals from the source space with enough sensitivity resolution. Figure 4 shows that the range of the HSV of our cap is $\sim 2-23\,mm^3$. Since there is no reference to compare but the reports using spheres fitted to the human head, we shall appeal to the aforementioned 10-scale rule. For separations equivalent to a human 129 channel cap there are reports of $HSV = 2.8 cm^3$ (Malmivuo et al., 1997) or $HSV \sim 6 - 8 cm^3$ (Ferree et al., 2001). For low density montages, such as the 19 channel cap, $HSV \sim 22 - 37 cm^3$ (Ferree et al., 2001). According to this, we could conclude that our rat EEG mini-cap has a variable sensitivity resolution spanning from low to high density human EEG caps, depending on the measurement site.

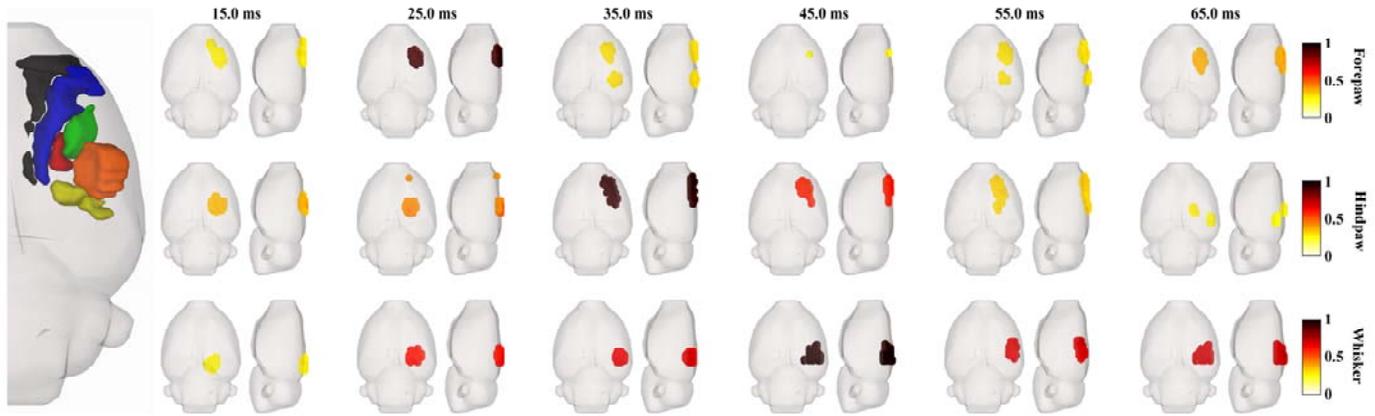

Fig. 14. **ESI in the volumetric gray matter model for forepaw, hindpaw and whisker SEPs, estimated using the template LF matrix.** The absolute value of the average of the inverse solutions with group constraints using SPM-LOR is shown. The light gray surface represents the brain. To the left, the green, red, blue, orange, yellow and black surfaces (these two appear only in the bottom row) represent S1FL, S1HL, M1, SLBF, Pt and M2, respectively. Only the 99 percentile of the values of the inverse solutions are shown. Besides, values below 25% of the total maxima across surface vertices and time points were also discarded to avoid spurious inverse solutions at instants of no expected ESI activity. The colour bars linearly map the values from the lowest to the highest value across all vertices and time instants. Notice that forepaw ESI activations in this figure are different from to the one in Figure 13. This is because the results in this figure is an average after a grouped estimation and the result in Figure 13 is the estimation of an individual case.

Alternatively, the sensitivity resolution can be also reported independent of the scale of the model, i.e. in terms of the volume fractions relative to the volume of the brain. The rat brain volume is .. (Valdés-Hernández et al., 2011), while the spherical brain volume fitted to a human head is $683 cm^3$ (Malmivuo et al., 1997). Following again the 10-scale rule, given two HSV with the same values, but one in $mm^3$ for the rat and the other in $cm^3$ for the human head, the volume fraction relative to the source space of the former is about 2.8 times smaller than that of the latter and so the sensitivity resolution of the former is 2.8 better than the latter.

Using volume fractions as measures of the relative sensitivity resolution tend to overestimate the resolution of the mini-cap, as compared to that of the human spherical model. This is because the elongated shape of the rat head model, which provides a large contribution to the source space model. We propose a different relative measure, more independent of this anisotropic shape. This is the ratio of the depth of the HSV region to the IS distance of the brain. The average maximum depth of the HSV region in our study is $\sim 1.5 mm$, thus this ratio is $\sim 1.5/11 \approx 0.14$. A similar value is obtained for the spherical human head, i.e. $\sim 2.5/2 \times 8 \approx 0.156$. For this math, we've used the maximum depth of $2.5 cm$ reported by Ferree et al. (2001) and the diameter of the sphere (radius = 8cm) as the IS distance.

A similar comparative analysis can be applied to the CRLB results. An exhaustive evaluation of the CRLB for different electrode configurations in the spherical model was described by Mosher et al. (1993). Among these, the so-called high density configuration of 37 electrodes, all grouped in the top of the sphere, is the one that mostly resembles our rat EEG mini-cap. Like with our mini-cap, the way in which the electrodes are spatially distributed furnishes a laminar configuration of increasing RMS values for increasing depth. We suggest that the reader compare our results for the noise only model, presented in Figure 6B, with the Figure 10 of that paper. Our value $f = 250 \Omega/m$ is 6 times higher than that used in Mosher et al. (1993). Their suggestions were $\sigma = 0.4 \mu V$, $Q = 10 nAm$ and thus $f = 40 \Omega/m$. If we use this value in our noise only model, the RMS must be multiplied by 6. This yields RMS similar to those in the aforementioned electrode configuration presented by Mosher et al., (1993), if we appeal to the 10-scale rule for comparisons.

### B. Using sample estimators of the covariance matrix

The assessment of the CRLB for EEG forward models was firstly used to investigate the performance of MUSIC and the Maximum Likelihood estimator (Stoica and Nehorai, 1989). Also, an exhaustive evaluation of the CRLB for different electrode configurations in the spherical model was described by Mosher et al. (1993). Yet, these initial approaches considered that the model is correct and only corrupted by unknown noise at the sensors. More recent papers investigated the effect of errors in the forward model on the CRLB, as well as on the source estimation error. For the general forward equation (B1), these model errors are as many as the variables that determine the LF matrix. For example, uncertainty of the skull conductivity was considered by Plis et al. (2007). They also evaluated the CRLB when this conductivity is also an unknown parameter to be estimated. On the other hand, the effect of incorrect electrode position was tackled by Beltrachini et al. (2011). Uncertainty of the parameters that characterize the surfaces defining the conductivity compartments was the subject of the CRLB analysis in von Ellenrieder et al. (2006). In those papers, the uncertainties (or errors) of the parameters that determine the LF matrix were theoretically modelled and propagated to the covariance matrix through the forward model. On the contrary, in the present paper, we use the actual variability of the LF matrices present in the sample to directly estimate the data covariance matrix. This approach could be biased by the lack of enough samples to provide an asymptotic estimate. However, the CRLB results in the aforementioned papers could also be biased because the parameters and functions chosen to model the uncertainties could be wrong. We think that a correct strategy might be estimating the data covariance matrix by characterizing the variability of either the LF matrix or the parameters that determine it, using a large enough sample. For too small samples, shrinkage estimators of the covariance matrix must be employed (Chen et al., 2009; Ledoit and Wolf, 2003; Wiesel and Hero, 2010). In the latter case, the way the estimated uncertainty of the parameters is propagated to the covariance matrix must be tackled carefully, for the variability might not be small enough to consider a linear relation between the LF matrix and the parameters, as was done by von Ellenrieder et al., (2006).

Table 2. **Average and standard deviation across the structures, of either each rat cortical structure or conglomerate of structures, of the RMS, RLE and Z**. Since the first three single structures are not conglomerates, the RLE is calculated dividing the results of the first two columns by the average linear size of the structure. The "Rhinal and Insular" conglomerate is highlighted because tis RLE > 50%.

| Cortical Region | RMS (mm) | | RLE | | Z | |
|---|---|---|---|---|---|---|
| | Noise Only | Model Errors | Noise Only | Model Errors | Noise Only | Model Errors |
| **S1FL** | 0.62 ± 0.18 | 0.95 ± 0.21 | 12.2 ± 3.6 | 18.9 ± 4.1 | 10.82 ± 2.26 | 6.48 ± 0.82 |
| **S1HL** | 0.39 ± 0.10 | 0.72 ± 0.17 | 10.0 ± 2.7 | 18.5 ± 4.3 | 15.74 ± 2.66 | 8.18 ± 0.54 |
| **S1BF** | 0.83 ± 0.30 | 1.21 ± 0.33 | 13.1 ± 4.8 | 19.2 ± 5.3 | 10.26 ± 2.37 | 6.35 ± 0.93 |
| **Cingulate** | 0.79 ± 0.34 | 1.11 ± 0.30 | 15.7 ± 4.9 | 22.8 ± 8.3 | 9.71 ± 3.77 | 6.25 ± 1.41 |
| **Motor** | 0.93 ± 0.66 | 1.18 ± 0.60 | 10.2 ± 2.7 | 12.7 ± 2.0 | 9.49 ± 5.68 | 5.66 ± 2.18 |
| **Parietal Association** | 0.33 ± 0.07 | 0.67 ± 0.13 | 10.3 ± 2.1 | 21.3 ± 4.9 | 18.16 ± 2.75 | 9.03 ± 0.54 |
| **Retrosplenial** | 0.46 ± 0.20 | 0.86 ± 0.25 | 7.5 ± 3.3 | 14.6 ± 6.8 | 15.45 ± 3.97 | 8.06 ± 1.25 |
| **Auditory** | 0.43 ± 0.16 | 0.78 ± 0.21 | 7.7 ± 1.7 | 14.2 ± 1.9 | 14.33 ± 4.08 | 7.41 ± 1.48 |
| **Primary Somatosensory** | 0.94 ± 0.50 | 1.29 ± 0.50 | 20.6 ± 12.2 | 29.0 ± 13.0 | 10.03 ± 3.64 | 6.15 ± 1.41 |
| **Secondary Somatosensory** | 0.49 ± 0.13 | 0.85 ± 0.18 | 26.0 ± 21.0 | 40.6 ± 29.6 | 13.29 ± 2.89 | 7.41 ± 0.98 |
| **Parietal** | 1.61 ± 0.66 | 1.95 ± 0.63 | 34.1 ± 17.3 | 40.4 ± 17.8 | 5.40 ± 2.54 | 4.10 ± 1.26 |
| **Visual** | 2.18 ± 0.34 | 2.48 ± 0.32 | 37.3 ± 5.9 | 42.5 ± 5.5 | 3.95 ± 1.21 | 3.59 ± 0.85 |
| **Rhinal and Insular** | 3.86 ± 1.82 | 4.03 ± 1.74 | 74.4 ± 40.6 | 77.7 ± 40.3 | 2.24 ± 1.63 | 2.15 ± 1.12 |

### C. Further theoretical considerations

The sensitivity of the electrodes could also be investigated using different approaches like (Väisänen et al., 2008). In that paper, the so-called region-of-interest sensitivity ratio (ROISR) is defined to quantify how well the EEG sensitivity is concentrated in a region of interest in the brain. Interestingly, the ROISR is correlated with the Signal-to-Noise ratio (SNR). The ROISR provides a measure of EEG measurement specificity. The application of this method to realistic source models, in the way we proceeded in this paper, could be an interesting subject for future studies.

On the other hand, we envisage from the Appendix B that there must be an improvement in the RMS when having multiple time samples. Intuitively this implies a "bigger" Fisher information matrix and thus a lower CRLB. For mathematical simplicity, let us impose the same temporal behavior $Q(t) = Qc(t)$ to all dipoles, where $c(t)$ is a fraction between $[0,1]$. Let us also deal with the noise only case. In Eqs. (B6), (B10) and (11) for calculating the RMS, the f factor that scales the RMS is now corrected: $f \rightarrow f / \sqrt{\sum_{t=1}^{N_T} c^2(t)}$. Suppose that the dipole temporal behavior is reflected in the EEG signal. Let's then use the temporal behavior of the channel with the maximum value in the EEG-fMRI experiment (see Figure 11). For the estimation interval 5-70 ms, $\sqrt{\sum_{t=5ms}^{70ms} c^2(t)} = 3.75 > 1$, which means that $RMS \rightarrow RMS/3.75$. We stress that this sum actually might not be above 1. In this case, it was because the maximum of the signal, which occurs at 25ms, is within the time interval considered for estimation. The dependence of Z on $Q(t)$ is not so straightforward. However, the presence of multiple time points should also improve Z since it increases with the decrease of the RMS. We stress that we've only provided a rough intuition using the simplest case in a simplified situation. It was not our interest in this paper to investigate the possible effect of multiple time points on the full CRLB expressions. We propose the formulation and encourage the work of future theoretical studies.

### D. EEG source imaging of SEP in rats

Figures 12 and 13 show that the ESI activation in the EEG-fMRI experiment is appreciably different between eLORETA and SPM-LOR. While for the former the ESI activations at 15 ms, $25ms$ and $65ms$ are almost completely localized within the S1FL region, for the latter the ESI activations are shifted toward the central part of M1. Actually we cannot ascertain which method is outperforming the other since we don't know exactly where the ESI activity should actually be. If we follow the classical somatotopic criterion (Brecht et al., 2004; Chapin et al., 1987), the ESI activity should be always at S1FL, favoring the eLORETA. In this case, the displacement of the SPM-LOR solutions would be simply the consequence of the known bias in localization error of LORETA methods (Pascual-Marqui, 2007). However, recent functional studies using Voltage Sensitive Dye (VSD) revokes this anatomical constraint (Ghosh et al., 2009; Morales-Botello et al., 2012). Electrical excitation results not only in sensory but also in proprioceptive input due to feeble muscle and joint movements. On the other hand, we consider that the ESI activity correlates with the fMRI, then a small shift of the former toward M1 is possible, especially around 25ms and 65ms, as previous studies suggest (Franceschini et al., 2010, 2008). These reveal that the components of the SEPs that most correlate with diffuse optical imaging (DOI) of hemodynamic responses are N1 and P2, rather than P1.

Let us compare our ESI results for the stimulations of the paws with those reported using VSD in rats (Ghosh et al., 2009; Morales-Botello et al., 2012). VSD maps the neuronal activity of the supra-granular 2/3 layer over large spatial scales with very high temporal resolution. It is highly correlated with the local field potentials (LFP) (Ferezou et al., 2007; Petersen et al., 2003). In a VSD study, the instant of appearance of the first activity (called "amplitude latency") and the peak latency were found to be dependent on the intensity of the current applied to the paws (Morales-Botello et al., 2012). The mean and standard deviation across different rats at these latencies were higher for 0.6 mA than for 6 mA. We have not found experimental reports of the latencies for the 2-3 mA range we use in our experiments. Thus, we applied an inverse linear rule to provide a *guess* of the range in which our latencies might be. We stress that these *are not* real experimental values. We estimated that the contralateral amplitude latency and peak latency of the forepaw stimulation might be around $\sim 14 \pm 5ms$ and $\sim 24 \pm 6ms$, respectively. On the other hand, our forepaw ESI amplitude latency is 15ms for the EEG-fMRI experiment and $14ms$ for the other forepaw experiment, while the peak latencies are 25ms and 26ms, respectively. In the hindpaw stimulation, our estimated ranges are $\sim 19 \pm 4ms$ for the amplitude latency and $\sim 29 \pm 6ms$ for the peak latency; while our ESI amplitude and peak latencies are 14ms and 33ms, respectively. In *summary*, although the hindpaw amplitude latency of our experiment was low, the latencies are in general within the rough *guessed* ranges.

In our forepaw experiments, the spread and split of the ESI activity after the amplitude latency toward posterior or anterior regions is expected according to VSD results (Ghosh et al., 2009; Morales-Botello et al., 2012). Besides, the later spread toward motor regions in the contralateral and ipsilateral hemispheres of our EEG-fMRI ESI activations resembles the VSD images. The spread of contralateral VSD activity after the amplitude latency is more pronounced for the forepaw stimulation than for the hindpaw stimulation (Morales-Botello et al., 2012). This is consistent with the higher spread of ESI activation in our forepaw experiments, as compared with the hindpaw experiment (see Figure 14).

On the other hand, our results for the whisker stimulation are somewhat arguable and provocative. Based on VSD images and LFP recordings (Petersen et al., 2003), the first activation is expected to be localized within S1BF. During a later time interval, the VSD activity spreads toward S2, the contralateral and ipsilateral M1 and anteriorly to the Agranular Medial Motor cortex (Agm) (Ferezou et al., 2007; Matyas et al., 2010; Tsytsarev et al., 2010), which is the central and anterior part of the secondary motor cortex (M2) as defined in the Paxinos & Watson atlas[8]. This spatiotemporal behaviour of the activity is in agreement with Current Source Density (CSD) maps in layer 2/3 based on LFP recordings in both contralateral and ipsilateral S1, S2 and Agm (Mégevand et al., 2008). However, in our experiments the ESI activity appears at Pt and only moves toward the barrel field after 40ms. There might be some explanations to this clear disagreement. First, our experimental setup has several differences with that of the aforementioned studies, e.g. whole-whisker air puff vs. selected-whisker piezoelectric/tube, analgesia vs. anesthesia, scalp electrodes vs. subcutaneous/epicranial, and rats vs. mice. But nevertheless, we believe that the fundamental cause is the careless inclusion of multimodal stimulation. First, the use of a 10ms-duration air puff was visibly stimulating both the rat face and part of the body. The somatosensory representation of the body is the most posterior part of the somatosensory cortex, colliding with the parietal cortex (Brecht et al., 2004). This, combined with a possible localization bias of the LORETA, might be the reason of the ESI localization in Pt. But second, and probably more likely, the picopump used to create the air puffs produces a sharp sound (a click) that cannot be ignored since the rat is not anesthetized. Thus, there might be a simultaneous somatosensory and auditory stimulation. If so, the activation at the posterior parietal cortex and its surroundings is expected, since the parietal cortex is the epicenter of the multisensory integration in the rat (Lippert et al., 2013). Instead of discarding this data, given the inconsistency with classical results using other modalities, we decided to present the experiment to illustrate the scope of ESI in rats to deal with complex inquiries regarding global integration/processing of information across sensory and motor areas where consciousness is not compromised. This type of map of brain activity is not possible using other non-invasive modalities like fMRI, which do not have the temporal resolution needed to separate the different components of the sensorial responses.

---

[8] To avoid confusion: the Agm is considered by some authors as part of the primary motor cortex and M1 is called the Agranular Lateral Motor cortex (Agl) (Brecht et al., 2004). The whisker somatosensory motor region is suggested to be either at the Agm (Brecht et al., 2004) or in the boundary between Agm and Agl (Smith y Alloway, 2013).

In general, the significant ipsilateral ESI activity is difficult to detect when there is a high significant contralateral ESI activity. This is because, in that case, the ipsilateral is not above the 99% chosen for thresholding the inverse solutions. Note that, as suggested by VSD studies, the ipsilateral maximum is 40-50% of the contralateral one (Ghosh et al., 2009; Morales-Botello et al., 2012). Besides, averaging the inverse solutions of the second set of SEP experiments cancels the ipsilateral ESI activity that appears in the individual solutions. This is because its spatial and temporal variability is much higher than that of the contralateral ESI activity, consistent with VSD-based results in paw stimulations (Morales-Botello et al., 2012). On the other hand, the ipsilateral ESI activity is more pronounced in the EEG-fMRI experiment than in the individual solutions of the second set of SEP experiments. This might be a consequence of the different anesthesia used. In a previous study, inter-hemispheric BOLD correlations were lower when using medetomidine (Domitor, a sedative very similar to Dexdomitor) than when using α-chloralose (Williams et al., 2010). We used α-chloralose anesthesia in our EEG-fMRI experiment, while the dexdomitor was used in the second set of experiments.

In general, the 99 percentile threshold does not correctly detect the ESI activity, at a given instant, in different spatial regions with large differences in amplitude. However, decreasing this threshold introduces spurious inverse solutions. We consider that proper statistical confidence intervals, such as the desparsification methods (van de Geer et al., 2014; Zhang y Zhang, 2014), must be used in future studies to correctly estimate the significant activity. This is still an unsolved problem in ESI nowadays.

## VI. CONCLUSIONS

In this paper, we evaluated methodological and theoretical aspects related to the EEG forward modelling of rats. We demonstrated that a valid ESI can be achieved using non-invasive scalp EEG recordings in rats. We also demonstrated that the LF matrix built from the minimum deformation MRI template can be used as a surrogate of the individual LF matrix without drastically compromising the accuracy of ESI.

To this end, we first defined both BEM and FEM forward models for rats and calculated the LF matrices for the template head model and the models extracted from individual MRIs of a database. The electrodes of a 32-channel scalp EEG mini-cap (Riera et al., 2012) were identified in the models by means of a strategy based on landmarks defined in the mini-cap scaffold.

We evaluated the electrode sensitivity resolution of the mini-cap and the minimum theoretical separation of electrodes measuring non-redundant information. This was done using the HSV and FSV. Since the rat head model is very dissimilar from spherical models, we developed a methodology to evaluate the sensitivity of the electrodes at each ML and AP pair of electrodes of a 32-channel scalp EEG mini-cap. For this realistic geometry, we have detected that the curves of HSV/FSV versus lead separation do not saturate after a certain lead separation like those of spherical models. They actually fall, affected by the separation from the source space of one of the electrodes of the lead. This suggests that, for electrode pairs in the boundary of the mini-cap, the main contribution to the sensitivity comes from radial dipoles. In general, the sensitivity resolution is worse in the boundaries of the cap (especially in anterior and posterior regions) and worse for AP electrode pairs than for ML ones. We have also demonstrated that the rat EEG mini-cap does not measure redundant information. Besides, the limit for cap density in rat scalp EEG is lower in the boundaries of the scalp measurement region, and is probably worse beyond that. These extremes establish a density limit to the whole cap corresponding to about 3 mm and 3.5 mm electrode separation along the ML and AP directions, respectively, provided that cap density is constant to avoid source estimation biases. Therefore, for the current scalp measured region, 32 seems to be nearly the maximum recommendable number electrodes, though perhaps density could be slightly increased in the AP direction. In general, in order to increase the number of electrodes, the measurement region must be extended beyond the current limits. However this might not represent a significant improvement given the significant decrease in the sensitivity resolution and the consequent redefinition of the whole cap density.

By means of the Cramér-Rao theorem, we calculated the lower bounds of the error of any unbiased estimator of dipole location, orientation and amplitude for the model in Eq. (B1). To assess these bounds we incorporated uncertainty in the LF matrix to account for imprecisions in the electrode locations and mismatches in the tissue-limiting surfaces. This feigns the situation in which the template is used as a surrogate of the actual individual. In contrast to previous approaches (Beltrachini et al., 2011; von Ellenrieder et al., 2006), we estimated the data covariance matrix of the model using sample estimators using the individual LF matrices of the database. Specifically, the CRLB was evaluated using two measures: RMS and Z. The former is the root-mean-square of the CRLB of the localization error of the dipole. The latter is proposed in this paper for the first time and it is the ratio of the dipole amplitude to its CRLB root-mean-square, quantifying "detectability" or "estimability" of the dipoles.

According to the CRLB results, for SEP experiments, the RMS of a dipole located at a primary somatosensory structure is about 10-20% of the root-mean-square of the linear sizes of the region. Informally said, the probability that the estimated dipole is located within the structure is very high. On the other hand, two dipoles cannot be estimated if their separation is below 3-5 mm. This means that, in practice, two equivalent dipoles within the same anatomical structure of the Paxinos & Watson atlas cannot be estimated. We remark that we obtained this estimate with one dipole located in the cortex, close to the electrodes. The results might be worse if both dipoles are deeper. These results are only for one time instant. We envisage that for multiple time instants the RMS might be lower, as well as the effect of using the template LF matrix.

Our Cramér-Rao results also reveals that, for the typical range of noise level in ERP experiments, the use of the template with approximated electrode positions yields a slightly higher

RMS than using the actual individual LF matrix. For the single dipole, the average across all source points of the difference of the RMS between both methods is about 0.5 mm, which corresponds to about 3-4 voxels in the template MRI. Since CRLB are only necessary conditions for the accuracy of the estimated sources, we provide further evidence by showing that the correlation between the template LF matrix and the individual LF matrices is very high, for both the BEM and the FEM models.

Finally, we performed ESI in real SEP experiments using eLORETA and SPM-LOR to demonstrate the accuracy of the whole methodology. We first estimated the ESI activity of a joint EEG-fMRI experiment after forepaw electrical stimulation. The estimated activations using the individual and the template LF matrix were very alike, providing experimental face validity for the use of the template. The activations using eLORETA localized within S1FL and the activations using SPM-LOR localized within both S1FL and M1. Either result might be expected if we consider that the fMRI activation was shared by both anatomical areas, though predominantly by S1FL. For these data, the latencies were expected and the contralateral and ipsilateral spread resembled that reported using VSD. The second set of SEP experiments consisted on forepaw, hindpaw and whisker stimulation of four rats. The average solutions were shown and discussed. Although they lack the expected ipsilateral ESI activity reported in the literature, the contralateral main activation sites and latencies were correct for the paw stimulations.

Although our methodology and results were mainly based on the aforementioned EEG mini-cap, they are very likely valid for any rodent scalp EEG study, since mouse and rat heads are very similar in shape.

## Appendix A. Notation

Lowercase bold letter are column vectors, uppercase bold letters are matrices and non-bold letters are scalars, with the exception of the matrix-derived concatenation vectors defined with bold uppercase letters and an arrow, such as $\vec{J}$ and $\vec{V}$ (see Appendix B and Materials and Methods). The matrix $\mathbf{I}_N$ is the $N-th$ order identity matrix and $\mathbf{0}_{N \times M}$ is the $N \times M$ matrix of zero elements. The operator $\otimes$ is the Kronecker product. The selector vector $\mathbf{1}_p$ has elements $\left( \mathbf{1}_p \right)_q = \delta_{pq}$.

## Appendix B. Cramér-Rao Lower Bounds for the dipoles estimation error in the presence of noise and uncertainties of the LF matrix

The EEG forward model for time varying signals and sources is formulated as follows:

$$\mathbf{V}(\mathbf{J}, \mathbf{r}) = \mathbf{K}(\mathbf{r})\mathbf{J} + \mathbf{E}, \quad (B1)$$

where,

- $\mathbf{V} \in \mathbb{R}^{N_E \times N_T}$ is the potential, for $N_E$ sensors and $N_T$ time instants.

- $\mathbf{r} \in \mathbb{R}^{3N_D \times 1}$, with $\mathbf{r} = \left[ \mathbf{r}_1^T \; \ldots \; \mathbf{r}_{N_D}^T \right]^T$ stores the positions of the dipoles, i.e. $\mathbf{r}_j = \left[ r_{jx} \quad r_{jy} \quad r_{jz} \right]^T$ is the position of the $j-th$ dipole.

- $\mathbf{J} \in \mathbb{R}^{3N_D \times N_T}$ is the matrix of dipole values in all positions and instants, i.e. $\mathbf{J} = \left[ \mathbf{j}(1) \; \ldots \; \mathbf{j}(N_T) \right]$, $\mathbf{j}(t) = \left[ \mathbf{j}_1^T(t) \; \ldots \; \mathbf{j}_{N_D}^T(t) \right]^T$ and $\mathbf{j}_j(t) = \left[ j_{kx}(t) \quad j_{ky}(t) \quad j_{kz}(t) \right]^T$.

- $\mathbf{K}(\mathbf{r}) \in \mathbb{R}^{N_E \times 3N_D}$ is the LF matrix, which is only a function of the position of the dipoles. We can write $\mathbf{K}(\mathbf{r}) = \left[ \mathbf{K}(\mathbf{r}_1) \; \ldots \; \mathbf{K}(\mathbf{r}_{N_D}) \right]$ where $\mathbf{K}(\mathbf{r}_j) \in \mathbb{R}^{N_E \times 3}$ is the contribution of the $j-th$ dipole[9]. We are not considering moving dipoles, thus the LF matrix is independent of time.

- $\mathbf{E} = \left[ \boldsymbol{\varepsilon}(1) \; \ldots \; \boldsymbol{\varepsilon}(N_T) \right] \in \mathbb{R}^{N_E \times N_T}$, is a multivariate Gaussian white noise in the sensors, i.e. $\boldsymbol{\varepsilon}(t) \sim N\left( \mathbf{0}, \sigma^2 \mathbf{I}_{N_E} \right)$ and $\mathrm{Cov}\left\{ \boldsymbol{\varepsilon}(t_1), \boldsymbol{\varepsilon}(t_2) \right\} = \delta_{t_1, t_2}$.

The error due to model uncertainty is:

$$\Delta\mathbf{K}(\mathbf{r}) = \mathbf{K}(\mathbf{r}) - \mathbf{K}^{(0)}(\mathbf{r}) \quad (B2)$$

where is the individual LF matrix and $\mathbf{K}^{(0)}$ is the template LF matrix. Each element of $\Delta\mathbf{K}$ is modelled as zero-mean Gaussian and so is the voltage vector. In order to define the likelihood given all dipole positions at all instants we define the vectors $\vec{\mathbf{V}} = \left[ \mathbf{v}^T(1) \; \ldots \; \mathbf{v}^T(N_T) \right]^T$ and $\vec{\mathbf{J}} = \left[ \mathbf{j}^T(1) \; \ldots \; \mathbf{j}^T(N_T) \right]^T$ [10] and pose:

$$\vec{\mathbf{V}} \sim N\left( \mathbf{G}_0\vec{\mathbf{J}}, \mathbf{C}_V \right)$$
$$\mathbf{C}_V\left( \vec{\mathbf{J}}, \mathbf{R} \right) = E\left\{ \Delta\mathbf{G}(\mathbf{R})\vec{\mathbf{J}}\vec{\mathbf{J}}^T\left( \Delta\mathbf{G}(\mathbf{R}) \right)^T \right\} + \sigma^2\mathbf{I}_{N_E N_T} \quad (B3)$$
$$\mathbf{G}_0 = \mathbf{I}_{N_T} \otimes \mathbf{K}_0$$
$$\Delta\mathbf{G}_0 = \mathbf{I}_{N_T} \otimes \Delta\mathbf{K}$$

For a vector of parameters to be estimated, $\boldsymbol{\theta}$, the Cramér-Rao theorem states that the covariance of the errors of the estimators $\hat{\boldsymbol{\theta}}$ cannot be lower than the inverse of the Fisher information matrix[11]:

$$E\left\{ \left( \boldsymbol{\theta} - \hat{\boldsymbol{\theta}} \right)\left( \boldsymbol{\theta} - \hat{\boldsymbol{\theta}} \right)^T \right\} \geq \mathbf{F}^{-1}$$
$$\mathbf{F} = E\left\{ \left[ \frac{\partial}{\partial\boldsymbol{\theta}} \log p\left( \vec{\mathbf{V}} \mid \boldsymbol{\theta} \right) \right]\left[ \frac{\partial}{\partial\boldsymbol{\theta}} \log p\left( \vec{\mathbf{V}} \mid \boldsymbol{\theta} \right) \right]^T \right\} \quad (B4)$$

For Gaussian likelihoods the Fisher information matrix is:

---

[9] We have adopted a convenient notation that explicitly expresses the dependency of the LF matrix on the position, but note that, according to the notation used in Eq. (13), $\mathbf{K}(\mathbf{r}_j) \equiv \mathbf{K}_j$.

[10] Note the first two equations in (B3) are actually the general form accounting for time varying and time auto-correlated LF matrices, for which case the matrices $\mathbf{G}_0$ and $\Delta\mathbf{G}$ are full matrices.

[11] Matrix $\mathbf{A}$ "no lower" than $\mathbf{B}$ means that $\mathbf{A} - \mathbf{B}$ is a semi-positive matrix.

$$F_{pq} = \left\{ \frac{\partial \vec{V}}{\partial \theta_p} \mathbf{C}_V^{-1} \frac{\partial \vec{V}}{\partial \theta_q} + \frac{1}{2} \mathrm{tr} \left\{ \mathbf{C}_V^{-1} \frac{\partial \mathbf{C}_V}{\partial \theta_p} \mathbf{C}_V^{-1} \frac{\partial \mathbf{C}_V}{\partial \theta_q} \right\} \right\} \ (B5)$$

Thus, if $\boldsymbol{\theta} = \begin{bmatrix} \sigma^2 & \vec{\mathbf{J}}^T & \mathbf{r}^T \end{bmatrix}^T$ the Fisher information matrix is then:

$$\mathbf{F} = \begin{bmatrix} \dfrac{N_E N_T}{2\sigma^2} & 0 & 0 \\ 0 & \mathbf{F}_{jj} & \mathbf{F}_{rj}^T \\ 0 & \mathbf{F}_{rj} & \mathbf{F}_{rr} \end{bmatrix} \quad \begin{aligned} \mathbf{F}_{jj} &= \mathbf{G}_0^T \mathbf{C}_V^{-1} \mathbf{G}_0 + \mathbf{F}_{jj}^{(2)} \\ \mathbf{F}_{rj} &= (\mathbf{DX})^T \mathbf{C}_V^{-1} \mathbf{G}_0 + \mathbf{F}_{rj}^{(2)} \\ \mathbf{F}_{rr} &= (\mathbf{DX})^T \mathbf{C}_V^{-1} (\mathbf{DX}) + \mathbf{F}_{rr}^{(2)} \end{aligned} \quad ,$$

(B6)

where $\mathbf{X} = \begin{bmatrix} \mathbf{I}_3 \otimes \mathbf{j}_1 & 0 & 0 \\ 0 & \ddots & 0 \\ 0 & 0 & \mathbf{I}_3 \otimes \mathbf{j}_{N_D} \end{bmatrix}$, $\mathbf{D} = \begin{bmatrix} \mathbf{d}(\mathbf{r}_1) & \dots & \mathbf{d}(\mathbf{r}_{N_d}) \end{bmatrix}$

stores the derivatives of the LF matrix with respect to the positions, being $\mathbf{d}(\mathbf{r}_k) \equiv \begin{bmatrix} \dfrac{\partial \mathbf{K}_0(\mathbf{r}_k)}{\partial r_{xk}}, & \dfrac{\partial \mathbf{K}_0(\mathbf{r}_k)}{\partial r_{yk}} & \dfrac{\partial \mathbf{K}_0(\mathbf{r}_k)}{\partial r_{zk}} \end{bmatrix}$.

For $\Delta\mathbf{K} = 0$ the first terms of the sub-matrices in (B6) are the same than in [Mosher et al. (1993)](#) while the second terms become zero. The latter are:

$$\left(\mathbf{F}_{jj}^{(2)}\right)_{pq} = \frac{1}{2} tr \left\{ \mathbf{C}_V^{-1} \frac{\partial \mathbf{C}_V}{\partial j_p} \mathbf{C}_V^{-1} \frac{\partial \mathbf{C}_V}{\partial j_q} \right\}$$

$$\left(\mathbf{F}_{rj}^{(2)}\right)_{pq} = \frac{1}{2} tr \left\{ \mathbf{C}_V^{-1} \frac{\partial \mathbf{C}_V}{\partial r_p} \mathbf{C}_V^{-1} \frac{\partial \mathbf{C}_V}{\partial j_q} \right\} \quad (B7)$$

$$\left(\mathbf{F}_{rr}^{(2)}\right)_{pq} = \frac{1}{2} tr \left\{ \mathbf{C}_V^{-1} \frac{\partial \mathbf{C}_V}{\partial r_p} \mathbf{C}_V^{-1} \frac{\partial \mathbf{C}_V}{\partial r_q} \right\}$$

$$\frac{\partial \mathbf{C}_V}{\partial q_p} = E\left\{ \mathbf{A}_p + \mathbf{A}_p^T \right\}$$

$$p = 3N_G(t-1) + 3(k-1) + c \qquad (B8)$$

$$t = 1 \dots N_T \quad k = 1 \dots N_G \quad c = 1,2,3$$

$$\frac{\partial \mathbf{C}_V}{\partial r_p} = E\left\{ \mathbf{B}_p + \mathbf{B}_p^T \right\}$$

$$p = 3(k-1) + c \qquad , (B9)$$

$$k = 1 \dots N_G \quad c = 1,2,3$$

where $\mathbf{A}_p = \Delta \mathbf{G} \mathbf{1}_p (\Delta \mathbf{GJ})^T$, $\mathbf{B}_p = (\Delta \mathbf{DX}) \mathbf{1}_p (\Delta \mathbf{G}\vec{\mathbf{J}})^T$, and like above, the derivatives of the LF matrix variations are $\Delta \mathbf{D} = \begin{bmatrix} \Delta \mathbf{d}(\mathbf{r}_1) & \dots & \Delta \mathbf{d}(\mathbf{r}_{N_d}) \end{bmatrix}$ and $\Delta \mathbf{d}(\mathbf{r}_k) \equiv \begin{bmatrix} \dfrac{\partial \Delta \mathbf{K}(\mathbf{r}_k)}{\partial r_{xk}}, & \dfrac{\partial \Delta \mathbf{K}(\mathbf{r}_k)}{\partial r_{yk}} & \dfrac{\partial \Delta \mathbf{K}(\mathbf{r}_k)}{\partial r_{zk}} \end{bmatrix}$. The dependence of the LF matrix on dipole positions is smooth enough to approximate their values as piece-wise linear functions with a high accuracy. Therefore we use finite-difference operators for calculating $\mathbf{D}$ and $\Delta\mathbf{D}$.

From Eq. (B6), the Cramér-Rao Lower Bound (CRLB) for the location of the dipoles is:

$$\mathbf{CRLB}(\mathbf{r}) = \left( \mathbf{F}_{rr} - \mathbf{F}_{rj}\mathbf{F}_{jj}^{-1}\mathbf{F}_{rj}^T \right)^{-1} \ (B10)$$

And the CRLB for the values of the current dipoles is:

$$\mathbf{CRLB}(\vec{\mathbf{J}}) = \left( \mathbf{F}_{jj} - \mathbf{F}_{rj}^T\mathbf{F}_{rr}^{-1}\mathbf{F}_{rj} \right)^{-1} \ (B11)$$

## ACKNOWLEDGMENT

We thank Lloyd Smith for his help on typos and English grammar; Dr. Eduardo Martinez Montes for his revision and suggestions and Joe Michel Lopez Inguanso for some useful searches in the Internet. We also thank Professor Pedro A. Valdés-Sosa for useful advices in the organization of the paper and the presentation of the main ideas.

## REFERENCES

Aghajani, H., Zahedi, E., Jalili, M., Keikhosravi, A., Vahdat, B.V., 2013. Diagnosis of early Alzheimer's disease based on EEG source localization and a standardized realistic head model. IEEE J. Biomed. Heal. Informatics 17, 1039-1045. doi:10.1109/JBHI.2013.2253326

Ashburner, J., Friston, K.J., 2005. Unified segmentation. Neuroimage 26, 839-51. doi:10.1016/j.neuroimage.2005.02.018

Ashburner, J., Friston, K.J., 2000. Voxel-based morphometry--the methods. Neuroimage 11, 805-821. doi:10.1006/nimg.2000.0582

Bae, J., Deshmukh, A., Song, Y., Riera, J., 2015. Brain Source Imaging in Preclinical Rat Models of Focal Epilepsy using High- Resolution EEG Recordings 1-12. doi:10.3791/52700

Beltrachini, L., von Ellenrieder, N., Muravchik, C.H., 2011. General bounds for electrode mislocation on the EEG inverse problem. Comput. Methods Programs Biomed. 103, 1-9. doi:10.1016/j.cmpb.2010.05.008

Bharadwaj, S., Gofton, T.E., 2015. EEG as a surrogate to brain imaging for diagnosing stroke in morbidly obese patients. J. Neurosurg. Anesthesiol. 27, 77-8. doi:10.1097/ANA.0000000000000074

Brecht, M., Krauss, A., Muhammad, S., Sinai-Esfahani, L., Bellanca, S., Margrie, T.W., 2004. Organization of rat vibrissa motor cortex and adjacent areas according to cytoarchitectonics, microstimulation, and intracellular stimulation of identified cells. J. Comp. Neurol. 479, 360-373. doi:10.1002/cne.20306

Chapin, J.K., Sadeq, M., Guise, J.L., 1987. Corticocortical connections within the primary somatosensory cortex of the rat. J. Comp. Neurol. 263, 326-46. doi:10.1002/cne.902630303

Chen, Y., Wiesel, A., Hero, A.O., 2011. Robust shrinkage estimation of high-dimensional covariance matrices. IEEE Trans. Signal Process. 59, 4097-4107. doi:10.1109/TSP.2011.2138698

Choi, J.H., Koch, K.P., Poppendieck, W., Lee, M., Shin, H.-S., 2010. High Resolution Electroencephalography in Freely Moving Mice. J. Neurophysiol. 104, 1825-1834. doi:10.1152/jn.00188.2010

Coben, R., Mohammad-Rezazadeh, I., Cannon, R.L., 2014. Using quantitative and analytic EEG methods in the understanding of connectivity in autism spectrum disorders: a theory of mixed over- and under-connectivity. Front. Hum. Neurosci. 8, 45. doi:10.3389/fnhum.2014.00045

Darvas, F., Ermer, J.J., Mosher, J.C., Leahy, R.M., 2006. Generic head models for atlas-based EEG source analysis. Hum. Brain Mapp. 27, 129-143. doi:10.1002/hbm.20171

Do Carmo, S., Cuello, A., 2013. Modeling Alzheimer's disease in transgenic rats. Mol. Neurodegener. 8, 37. doi:10.1186/1750-1326-8-37

Ferezou, I., Haiss, F., Gentet, L.J., Aronoff, R., Weber, B., Petersen, C.C.H., 2007. Spatiotemporal Dynamics of Cortical Sensorimotor Integration in Behaving Mice. Neuron 56, 907-923. doi:10.1016/j.neuron.2007.10.007

Ferree, T.C., Clay, M.T., Tucker, D.M., 2001. The spatial resolution of scalp EEG. Neurocomputing 38-40, 1209-1216. doi:10.1016/S0925-2312(01)00568-9

Franceschini, M.A., Nissilä, I., Wu, W., Diamond, S.G., Bonmassar, G., Boas, D. a., 2008. Coupling between somatosensory evoked potentials and hemodynamic response in the rat. Neuroimage 41, 189-203. doi:10.1016/j.neuroimage.2008.02.061

Franceschini, M.A., Radhakrishnan, H., Thakur, K., Wu, W., Ruvinskaya, S., Carp, S., Boas, D.A., 2010. The effect of different anesthetics on neurovascular coupling. Neuroimage 51, 1367-1377. doi:10.1016/j.neuroimage.2010.03.060

Gevins, a, Brickett, P., Costales, B., Le, J., Reutter, B., 1990. Beyond topographic mapping: towards functional-anatomical imaging with 124-channel EEGs and 3-D MRIs. Brain Topogr. 3, 53-64. doi:10.1007/BF01128862

Ghosh, A., Sydekum, E., Haiss, F., Peduzzi, S., Zorner, B., Schneider, R., Baltes, C., Rudin, M., Weber, B., Schwab, M.E., 2009. Functional and Anatomical Reorganization of the Sensory-Motor Cortex after Incomplete Spinal Cord Injury in Adult Rats. J. Neurosci. 29, 12210-12219. doi:10.1523/JNEUROSCI.1828-09.2009

Harmony, T., Fernández-Bouzas, A., Marosi, E., Fernández, T., Valdés, P., Bosch, J., Riera, J., Bernal, J., Rodríguez, M., Reyes, A., 1995. Frequency source analysis in patients with brain lesions. Brain Topogr. 8, 109-117.

Hughes, J.R., Ph, D., John, E.R., 1999. Conventional and Quantitative Electroencephalography in Psychiatry. J. Neuropsychiatry Clin. Neurosci. 11, 190-208.

Kaiboriboon, K., Lüders, H.O., Hamaneh, M., Turnbull, J., Lhatoo, S.D., 2012. EEG source imaging in epilepsy—practicalities and pitfalls. Nat. Rev. Neurol. 8, 498-507. doi:10.1038/nrneurol.2012.150

Kay, S.M., 1993. Fundamentals of Statistical Signal Processing: Estimation Theory, ser. Signal Processing.

Ledoit, O., Wolf, M., 2004. Honey, I Shrunk the Sample Covariance Matrix. J. Portf. Manag. 30, 110-119. doi:10.3905/jpm.2004.110

Lee, M., Kim, D., Shin, H., Sung, H., Choi, J.H., 2011. High-density EEG Recordings of the Freely Moving Mice using Polyimide-based Microelectrode. J. Vis. Exp. 2-5. doi:10.3791/2562

Lippert, M.T., Takagaki, K., Kayser, C., Ohl, F.W., 2013. Asymmetric Multisensory Interactions of Visual and Somatosensory Responses in a Region of the Rat Parietal

Cortex. PLoS One 8. doi:10.1371/journal.pone.0063631

Litvak, V., Friston, K., 2008. Electromagnetic source reconstruction for group studies. Neuroimage 42, 1490-1498. doi:10.1016/j.neuroimage.2008.06.022

Löscher, W., 2011. Critical review of current animal models of seizures and epilepsy used in the discovery and development of new antiepileptic drugs. Seizure 20, 359-368. doi:10.1016/j.seizure.2011.01.003

MacRae, I., 2011. Preclinical stroke research - Advantages and disadvantages of the most common rodent models of focal ischaemia. Br. J. Pharmacol. 164, 1062-1078. doi:10.1111/j.1476-5381.2011.01398.x

Malmivuo, J., Suihko, V., Eskola, H., 1997. Sensitivity distributions of EEG and MEG measurements. IEEE Trans. Biomed. Eng. 44, 196-208. doi:10.1109/10.554766

Malmivuo, J.A., Suihko, V.E., 2004. Effect of skull resistivity on the spatial resolutions of EEG and MEG. IEEE Trans. Biomed. Eng. 51, 1276-1280. doi:10.1109/TBME.2004.827255

Mattout, J., Henson, R.N., Friston, K.J., 2007. Canonical source reconstruction for MEG. Comput. Intel. Neurosci. 2007, 67613.

Matyas, F., Sreenivasan, V., Marbach, F., Wacongne, C., Barsy, B., Mateo, C., Aronoff, R., Petersen, C.C.H., 2010. Motor control by sensory cortex. Science 330, 1240-1243. doi:10.1126/science.1195797

Mégevand, P., Quairiaux, C., Lascano, A.M., Kiss, J.Z., Michel, C.M., 2008. A mouse model for studying large-scale neuronal networks using EEG mapping techniques. Neuroimage 42, 591-602. doi:10.1016/j.neuroimage.2008.05.016

Morales-Botello, M.L., Aguilar, J., Foffani, G., 2012. Imaging the Spatio-Temporal Dynamics of Supragranular Activity in the Rat Somatosensory Cortex in Response to Stimulation of the Paws. PLoS One 7, e40174. doi:10.1371/journal.pone.0040174

Mosher, J.C., Spencer, M.E., Leahy, R.M., Lewis, P.S., 1993. Error bounds for EEG and MEG dipole source localization. Electroencephalogr. Clin. Neurophysiol. 86, 303-321. doi:10.1016/0013-4694(93)90043-U

Pascual-Marqui, R.D., 2007. Discrete, 3D distributed, linear imaging methods of electric neuronal activity. Part 1: exact, zero error localization. Math. Phys. 1-16. doi:10.1186/1744-9081-4-27

Paxinos, G., Watson, C., 2007. The Rat Brain in Stereotaxic Coordinates, 6th ed. Elsevier Inc.

Petersen, C.C.H., Grinvald, A., Sakmann, B., 2003. Spatiotemporal dynamics of sensory responses in layer 2/3 of rat barrel cortex measured in vivo by voltage-sensitive dye imaging combined with whole-cell recordings and neuron reconstructions. J. Neurosci. 23, 1298-1309. doi:23/4/1298 [pii]

Plis, S.M., George, J.S., Jun, S.C., Ranken, D.M., Volegov, P.L., Schmidt, D.M., 2007. Probabilistic forward model for electroencephalography source analysis. Phys. Med. Biol. 52, 5309-5327. doi:10.1088/0031-9155/52/17/014

Quairiaux, C., Mégevand, P., Kiss, J.Z., Michel, C.M., 2011. Functional development of large-scale sensorimotor cortical networks in the brain. J. Neurosci. 31, 9574-

9584. doi:10.1523/JNEUROSCI.5995-10.2011

Riera, J.J., Fuentes, M.E., 1998. Electric lead field for a piecewise homogeneous volume conductor model of the head. IEEE Trans. Biomed. Eng. 45, 746-753. doi:10.1109/10.678609

Riera, J.J., Ogawa, T., Goto, T., Sumiyoshi, A., Nonaka, H., Evans, A., Miyakawa, H., Kawashima, R., 2012. Pitfalls in the dipolar model for the neocortical EEG sources. J. Neurophysiol. 108, 956-975. doi:10.1152/jn.00098.2011

Rodrak, S., Wongsawat, Y., 2013. EEG brain mapping and brain connectivity index for subtypes classification of attention deficit hyperactivity disorder children during the eye-opened period. Proc. Annu. Int. Conf. IEEE Eng. Med. Biol. Soc. EMBS 7400-7403. doi:10.1109/EMBC.2013.6611268

Rush, S., Driscoll, D.A., 1969. EEG electrode sensitivity--an application of reciprocity. IEEE Trans. Biomed. Eng. 16, 15-22. doi:10.1109/TBME.1969.4502598

Shevelkin, A. V., Ihenatu, C., Pletnikov, M. V., 2014. Pre-clinical models of neurodevelopmental disorders: Focus on the cerebellum. Rev. Neurosci. 25, 177-194. doi:10.1515/revneuro-2013-0049

Smith, J.B., Alloway, K.D., 2013. Rat whisker motor cortex is subdivided into sensory-input and motor-output areas. Front. Neural Circuits 7, 4. doi:10.3389/fncir.2013.00004

Stoica, P., Nehorai, A., 1989. Music, maximum likelihood, and cramer-rao bound. IEEE Trans. Acoust. Speech Signal Process. 37, 720-741.

Sumiyoshi, A., Riera, J.J., Ogawa, T., Kawashima, R., 2011. A mini-cap for simultaneous EEG and fMRI recording in rodents. Neuroimage 54, 1951-1965. doi:10.1016/j.neuroimage.2010.09.056

Tsytsarev, V., Pope, D., Pumbo, E., Yablonskii, A., Hofmann, M., 2010. Study of the cortical representation of whisker directional deflection using voltage-sensitive dye optical imaging. Neuroimage 53, 233-238. doi:10.1016/j.neuroimage.2010.06.022

Umeda, T., Takashima, N., Nakagawa, R., Maekawa, M., Ikegami, S., Yoshikawa, T., Kobayashi, K., Okanoya, K., Inokuchi, K., Osumi, N., 2010. Evaluation of Pax6 mutant rat as a model for autism. PLoS One 5, e15500. doi:10.1371/journal.pone.0015500

Väisänen, J., Väisänen, O., Malmivuo, J., Hyttinen, J., 2008. New method for analysing sensitivity distributions of electroencephalography measurements. Med. Biol. Comput. 46, 101-108. doi:10.1007/s11517-007-0303-x

Valdés-Hernández, P.A., Sumiyoshi, A., Nonaka, H., Haga, R., Aubert-Vázquez, E., Ogawa, T., Iturria-Medina, Y., Riera, J.J., Kawashima, R., 2011. An in vivo MRI template set for morphometry, tissue segmentation, and fMRI localization in rats. Front. Neuroinform. 5, 1. doi:10.3389/fninf.2011.00026

Valdés-Hernández, P.A., von Ellenrieder, N., Ojeda-Gonzalez, A., Kochen, S., Alemán-Gómez, Y., Muravchik, C., Valdés-Sosa, P.A., 2009. Approximate average head models for EEG source imaging. J. Neurosci. Methods 185, 125-132. doi:10.1016/j.jneumeth.2009.09.005

van de Geer, S., Bühlmann, P., Ritov, Y., Dezeure, R., 2014. On asymptotically optimal confidence regions and tests for high-dimensional models. Ann. Stat. 42, 1166-1202. doi:10.1214/14-AOS1221

von Ellenrieder, N., Muravchik, C.H., Nehorai, A., 2006. Effects of geometric head model perturbations on the EEG forward and inverse problems. IEEE Trans. Biomed. Eng. 53, 421-429. doi:10.1109/TBME.2005.869769

Vorwerk, J., Clerc, M., Burger, M., Wolters, C.H., 2012. Comparison of boundary element and finite element approaches to the EEG forward problem. Biomed. Tech. (Berl). 57 Suppl 1, 795-798. doi:10.1515/bmt-2012-4152

Wendel, K., Narra, N.G., Hannula, M., Kauppinen, P., Malmivuo, J., 2008. The influence of CSF on EEG sensitivity distributions of multilayered head models. IEEE Trans. Biomed. Eng. 55, 1454-1456. doi:10.1109/TBME.2007.912427

Wiesel, A., Hero, A.O., 2010. Robust Shrinkage Estimation of High-dimensional Covariance Matrices 189-192.

Williams, K. a., Magnuson, M., Majeed, W., LaConte, S.M., Peltier, S.J., Hu, X., Keilholz, S.D., 2010. Comparison of α-chloralose, medetomidine and isoflurane anesthesia for functional connectivity mapping in the rat. Magn. Reson. Imaging 28, 995-1003. doi:10.1016/j.mri.2010.03.007

Zhang, C.-H., Zhang, S.S., 2014. Confidence intervals for low dimensional parameters in high dimensional linear models. J. R. Stat. Soc. Ser. B (Statistical Methodol. 76, 217-242. doi:10.1111/rssb.12026